\documentclass[twoside]{article}
\usepackage{sprocl}
\usepackage{graphicx}
\usepackage{amssymb}

\newcommand{\dsp}{\displaystyle}
\newcommand{\ba}{\begin{eqnarray}}
\newcommand{\ea}{\end{eqnarray}}
\newcommand{\be}{\begin{equation}}
\newcommand{\ee}{\end{equation}}
\newcommand{\re}{\mbox{Re}}
\newcommand{\im}{\mbox{Im}}
\newcommand{\kob}{{\overline{K^0}}}
\newcommand{\epspeps}{{\varepsilon^\prime/\varepsilon}}

\newcommand{\hepph}[1]{{hep-ph/#1}}

\newcommand{\citetwo}[2]{\cite{#1,#2}}
\newcommand{\citethree}[3]{\cite{#1,#2,#3}}

\newcommand{\citefive}[5]{\cite{#1,#2,#3,#4,#5}}

\newcommand{\mycite}[1]{[\ref{biblio:#1}]}
\newcommand{\mycitetwo}[2]{[\ref{biblio:#1},\ref{biblio:#2}]}

\newcommand{\mybibitem}[1]
{\bibitem{#1} \label{biblio:#1}}

\begin{document}

\sloppy

\begin{titlepage}
\begin{flushright}
LU TP 02-12\\
hep-ph/0204068\\
April 2002
\end{flushright}
\vfill
\begin{center}
{\huge{\bf  QCD and Weak Interactions of Light Quarks}$^a$}
\end{center}

\vfill

\begin{center}
{\large\bf Johan~Bijnens}\\[1cm]
{\large Department of Theoretical Physics, Lund University,\\[0.5cm]
S\"olvegatan 14A, S22362 Lund, Sweden}
\end{center}

\vfill
\begin{center}
{\bf Abstract}
\end{center}
{This review contains an overview of strong interaction effects
in weak decays starting with a historical introduction. It contains a short
overview of semileptonic decays and their relevance for measuring CKM matrix
elements. The main part is devoted to the theoretical calculation
on nonleptonic matrix elements relevant for $K^0$-$\overline{K^0}$ mixing
and $K\to\pi\pi$ decays. It concludes with a short summary of
rare kaon decays.}
\vfill

\noindent $^a$
{\small
To be published in  'At the Frontier of Particle Physics / Handbook of QCD',
 edited by M. Shifman, Volume 4.}

\end{titlepage}


\pagestyle{myheadings}  
\markboth
{\small \em Handbook of QCD / Volume 4}
{\small \em QCD and Weak Interactions of Light Quarks}

\title{QCD AND WEAK INTERACTIONS OF LIGHT QUARKS}

\author{ JOHAN BIJNENS }

\address{Department of Theoretical Physics, Lund University,\\
S\"olvegatan 14A, S22362 Lund, Sweden}

\maketitle

\abstracts{This review contains an overview of strong interaction effects
in weak decays starting with a historical introduction. It contains a short
overview of semileptonic decays and their relevance for measuring CKM matrix
elements. The main part is devoted to the theoretical calculation
on nonleptonic matrix elements relevant for $K^0$-$\overline{K^0}$ mixing
and $K\to\pi\pi$ decays. It concludes with a short summary of
rare kaon decays.}

  \vspace{1cm}

\tableofcontents
\newpage

\section{Introduction}

This chapter gives an overview of the effects of Quantum ChromoDynamics (QCD)
in the weak interactions of light quarks. In this chapter I will discuss the
analytical knowledge existing at present for the main nonleptonic decays. 

Section~\ref{history} gives a short historical introduction to the area.
Since this is an overview of the theory involved, mainly theoretical
references are given. This is especially the case for the older papers.
The standard model is introduced briefly in Section \ref{standardmodel}.
After that we start with the first part of the influence of strong interactions
on the weak interactions, the influence of the strong interactions
on the semileptonic decays of light quarks in Section~\ref{semileptonic}.
Sections \ref{shortdistance} and \ref{longdistance} contain the main
parts of this introductory review, a description of the QCD effects
in $K^0$-$\kob$ mixing and $K\to\pi\pi$ decays. The last sections
contain a short overview of the pure Chiral Perturbation Theory
tests in nonleptonic kaon decays and of rare kaon decays.

\section{History}
\label{history}

The weak interaction was discovered in 1896 by Becquerel when he discovered
spontaneous radioactivity. This was the radioactive decay of uranium
and thus already in the beginning the study of the weak interaction was very
much entangled with the strong interaction. The study of radioactivity and
associated processes was a very active research area during much of the
beginning of the 20th century. The next large step towards a more fundamental
study of the weak interaction was taken in the 1930s when the neutron was
discovered and its $\beta$-decay studied in detail. After an initial
period of confusion since the proton and electron energies did not add up
to the total energy corresponding to the mass of the neutron,
Pauli suggested the neutrino as a way around this problem.
Fermi then incorporated this in the first full fledged theory of the
weak interaction, the famous Fermi four-fermion~\cite{Fermi} interaction. 
\be
\label{Fermi1}
{\cal L}_{\mbox{Fermi}} =
\frac{G_F}{\sqrt{2}}\; [\overline p \gamma_\mu\left(1-\gamma_5\right) n]\;
[\overline e \gamma^\mu\left(1-\gamma_5\right)\nu]\,.
\ee
Here I used the symbol for the particle itself also as a symbol
for its four-spinor.

Still, the weak interaction
was only known in a context mixing both strong and weak effects.
The first fully nonhadronic weak interaction came after world-war two
when the muon was discovered and one could study its $\beta$-decay.
The analogous Lagrangian to Eq.~(\ref{Fermi1}) was soon written down.
At that point T.D.~Lee and C.N.~Yang~\cite{LeeYang} realized that there was no
evidence that parity was conserved in the weak interaction. This quickly
led to a search for parity violation both in nuclear decays~\cite{CSWu}
by C.S.~Wu and collaborators and in the decay chain $\pi^+\to\mu^+\to e^+$
and its associated neutrinos.\cite{FriedmanTelegdi} Parity violation
was duly observed in both cases. Notice again that this was in processes
where the strong interaction plays an important role.

These experiments and others then led to the final form of the
Fermi Lagrangian given in Eq.~(\ref{Fermi1}) as proposed
by Sudarshak and Marshak~\cite{SudarshanMarshak} as well as
Feynman and Gell-Mann.\cite{FeynmanGell-Mann}

During the 1950s steadily more particles were discovered. In particular
the kaon together with new baryons. These particles provided
several puzzles. First, they seemed to be produced with strong interaction
probabilities but then were long-lived with lifetimes comparable
to typical weak interaction lifetimes of the pion and muon. This was solved
by the introduction of associated production, the strong interaction
does not violate a new quantum number, now called strangeness,
and can produce pairs, each particle with opposite strangeness.
This was introduced simultaneously by Pais~\cite{Pais}
and Gell-Mann.\cite{Gell-Mann} Further experimental study of these
particles revealed the second puzzle. There were two particles
with as far as could be measured, the same mass and the same
production mechanism. One of them was rather short-lived and decayed
to two pions while the other had a much longer lifetime and
decayed into three pions or semileptonically. This socalled
tau-theta puzzle was solved by Gell-Mann and Pais~\cite{Gell-MannPais}
who introduced what is now known as the $K_L$ and the $K_S$.

Further progress had to await the classification of the hadrons
into symmetry-multiplets, the $SU(3)_V$ of Ne'eman and
Gell-Mann~\cite{eightfold} building on earlier work by Sakata.\cite{Sakata}
Subsequently Cabibbo realized that the weak interactions of
the strange particles were very similar to those of the
nonstrange particles.\cite{Cabibbo} He proposed that the weak interactions
of hadrons occurred through a current which was a mixture of the
strange and non-strange currents with a mixing angle now universally
known as the Cabibbo angle. The symmetry group of the hadrons
led to the introduction of quarks~\cite{quarks} as a means of
organizing which $SU(3)_V$ multiplets actually were present in the hadronic
spectrum.

In the same time period the kaons provided another surprise. After the
discovery of parity violation and charge-conjugation violation
Lee and Yang had introduced $CP$ as a symmetry to save the Gell-Mann--Pais
mechanism to understand the tau-theta puzzle. The original suggestion
used charge-conjugation but the construction worked as well using the
combination of charge-conjugation and parity, $CP$. Measurements at
Brookhaven~\cite{CCFT} indicated that the long-lived state, the $K_L$,
did occasionally decay to two pions in the final state as well.
The last discrete symmetry thus fell as well. The remaining simple
discrete symmetry, the combination of $C$, $P$ and $T$, is
automatically satisfied in any local quantum field theory. At present
no such violations have been seen, but tests remain important given
that it tests such a basic aspect of our theories.
Since the $CP$-violation was small, explanations could be sought at
many scales, an early phenomenological analysis can be found
in Ref. ~\mycite{WuYang} but as Wolfenstein~\cite{Wolfenstein} showed,
the scale of the interaction involved in $CP$-violation could be much higher.
The latter became known as the superweak model for $CP$-violation.
At that time also some influential reviews were written discussing
weak interactions and in particular $C$, $P$, $T$, $CP$, $T$
and $CPT$ violation.\citethree{LeeWu1}{LeeWu2}{LeeWu3}

The standard model for the weak and electromagnetic interactions of leptons
was in the process of construction at the same time. 
Calculations with Fermi theory
for leptonic weak interactions led to infinities which were not
renormalizable. Alternatives based on Yang-Mills~\cite{Yang-Mills}
theories had been proposed by Glashow~\cite{Glashow} but struggled with
the problem of having massless gauge bosons. This was solved by 
the introduction of the
Higgs mechanism by Weinberg and Salam. The model
could be extended to include the weak interactions of hadrons by adding
quarks in doublets, similar to the way the leptons were included.
One problem this produced was that loop-diagrams provided a much too
high probability for the decay $K_L\to\mu^+\mu^-$ compared to
the experimental limits. These socalled flavour changing neutral currents
(FCNC) needed to be suppressed. The solution was found in
the Glashow-Iliopoulos-Maiani mechanism.\cite{GIM} A fourth quark, the charm
quark, was introduced beyond the up, down and strange quarks. This allowed
for a nice structure of the standard model with two generations of leptons
and quarks with the weak interactions acting very symmetrically between
quarks and leptons. A consequence of this, together with the extension
of the Cabibbo mixing to the two generation scenario, was that if all the
quark masses were equal, the dangerous loop contributions to FCNC processes
cancel. This cancellation is now generally known as the GIM mechanism.
An added advantage of this structure was that all anomalies of the
standard model gauge interactions cancel between the quark and the lepton
sector. This allowed a prediction of the charm quark
mass~\cite{GaillardLee1} which was soon borne out by experiment with
the discovery of the $J/\psi$ charm-anticharm state. 

In the mean time, as has been discussed in part 1 of these books,
QCD was formulated.\cite{QCD} The property of
asymptotic freedom~\cite{Asympt} was established which explained why
quarks at short distances could behave as free particles and at the
same time at large distances be confined inside hadrons. As discussed
elsewhere in these volumes the proof of this confinement or infrared
slavery is still one of the main open problems in the study of QCD.

The study of kaon decays still went on, and an already old problem,
the $\Delta I=1/2$ rule saw the first signs of a solution. It was
shown by Gaillard and Lee~\cite{GaillardLee2} and Altarelli and
Maiani~\cite{AltarelliMaiani} that the short-distance QCD part
of the nonleptonic weak decays provided already an enhancement of
the $\Delta I=1/2$ weak $\Delta S=1$ transition over the $\Delta I=3/2$
one. The ITEP group extended first the Gaillard-Lee analysis
for the charm mass,\cite{Vainshtein1} but then realized that in
addition to the effects that were included by Gaillard, Lee, Altarelli and
Maiani there was a new class of diagrams that only contributed to
the $\Delta I=1/2$ transition.\citetwo{Penguin1}{Penguin2} While, as we
will discuss in more detail later, the general class of these
contributions, now known generally as Penguin-diagrams, is the most likely main
cause of the $\Delta I=1/2$ rule, the short-distance part of them
provide only a small enhancement contrary to the original hope.
A description of the early history of Penguin diagrams, including the
origin of the name, can be found in the 1999 Sakurai Prize lecture  of
Vainshtein.\cite{Penguin3}

Penguin diagrams at short distances provide nevertheless a large amount
of physics. But here we need to return first to another (r)evolution that
happened in the mean time. The GIM mechanism had explained away the
troublesome FCNC effects but the origin of $CP$-violation was (and partly is)
still a mystery. The model of Wolfenstein explained it, but introduced new
physics that had no other predictions. 
Kobayashi and Maskawa~\cite{KobayashiMaskawa} realized that
the framework established by~Ref.~\mycite{GIM} and in particular how it
incorporated Cabibbo mixing could be extended to more than two
generations. The really new aspect this brings in is that $CP$-violation
could easily be produced at the weak scale and not at the much
higher superweak scale. In this scenario $CP$-violation comes from
the mixed quark-Higgs sector, the Yukawa sector from the standard model,
and is linked with the masses and mixings of the quarks.
Other mechanisms at the weak scale
also exist, in particular an extended Higgs sector provides
possibilities as well. This was pointed out by Weinberg.\cite{WeinbergCP}

The inclusion of the Kobayashi-Maskawa mechanism into the
calculations for weak decays was done by Gilman and
Wise~\citetwo{GilmanWise1}{GilmanWise2} which provided the prediction
that $\epspeps$ should be nonzero and of the order of $10^{-3}$ within
this picture. Guberina and Peccei~\cite{GuberinaPeccei} confirmed this.
This prediction spurred on the experimentalists and after
two generations of major experiments, NA48 at CERN and KTeV at Fermilab
have now determined this quantity and the qualitative prediction
that $CP$-violation at the weak scale exists is now confirmed.
Much stricter tests of this picture will happen at other kaon
experiments as well as at $B$ meson studies. The strong interaction
effects in the latter are described elsewhere in these volumes.

The $K^0$-$\kob$ mixing has QCD corrections and $CP$-violating contributions
as well. The calculations of these required a proper treatment
of box diagrams and inclusions of the effects of the $\Delta S=1$
interaction squared. This was accomplished at one-loop
by Gilman and Wise a few years later.\citetwo{GilmanWise3}{GilmanWise4}

That Penguins had more surprises in store was shown some years later
when it was realized that the enhancement originally expected on chiral grounds
for the Penguin diagrams~\citetwo{Penguin1}{Penguin2} was present,
not for the Penguin diagrams with gluonic intermediate states, but
for those with a photon.\cite{BijnensWise} This contribution
was also enhanced in its effects by the $\Delta I=1/2$ rule. 
This lowered the
expectation for $\epspeps$ but at that time not by very much.
A significant change came with the discovery of a large $B^0$-$\overline{B^0}$
mixing at DESY. This immediately led people to realize that the mass
of the top quark could be much larger than hitherto assumed.
Flynn and Randall~\cite{RandallFlynn} reanalyzed the electromagnetic
Penguin with a large top quark mass and included also $Z^0$ exchange
which could now be important as well. The final effect was that
the now rebaptized electroweak Penguins could have a very large contribution
that could even cancel the contribution to $\epspeps$\ from
gluonic Penguins. This story still continues at present and the
cancellation, though not complete, is one of the major impediments
to accurate theoretical predictions of $\epspeps$.

At that time, the short-distance part was fully analyzed at one-loop
but it became clear that higher precision would be needed.
The first calculation of two-loop effects was done in
Rome~\citetwo{twoloopold1}{twoloopold2} in 1981. The physics analyses
proceeded to only use the one-loop expressions since only for
these had the effects for all operators been calculated. The two-loop
calculation had also been performed in dimensional reduction,
a scheme known to have inconsistencies.
The value of $\Lambda_{\mbox{QCD}}$ continued to rise
from values of about 100~MeV to more than 300~MeV. A full calculation
of all operators at two loops thus became necessary, taking into account
all complexities of higher order QCD. This program was finally
accomplished by two independent groups. One in Munich around A.~Buras
and one in Rome around G.~Martinelli. These will be described
in Sec.~\ref{shortdistance}. The progress on evaluation of the
long-distance matrix elements since the original vacuum-insertion-approximation
from Gaillard and Lee can be found in Sec.~\ref{longdistance}.

\section{A Short Overview of the Standard Model}
\label{standardmodel}

The Standard Model Lagrangian has four parts:
\ba
{\cal L}_{SM} &=&
\dsp\underbrace{{\cal L}_H(\phi)}_{\mbox{Higgs}}
+
\underbrace{{\cal L}_G(W,Z,G)}_{\mbox{Gauge}}
\dsp+\underbrace{\sum_{\psi=\mbox{fermions}}
\bar\psi iD\hskip-0.7em/\hskip0.4em \psi}_{\mbox{gauge-fermion}}
\nonumber\\&&
+
\underbrace{\sum_{\psi,\psi^\prime=\mbox{fermions}}
g_{\psi\psi'}\bar\psi\phi\psi'}_{\mbox{Yukawa}}
\ea
The experimental tests on the various parts are at a
very different level:\\[0.5cm]
\begin{tabular}{ll}
gauge-fermion& Very well tested at LEP-1 and other precision\\
& measurements\\
Higgs& Hardly tested, LEP-1 and LEP-2 have made a start,\\& Tevatron
and LHC in the future\\
Gauge part& Well tested in QCD, partly in electroweak at LEP-2\\
Yukawa& The real testing ground for weak interactions and the\\& main source
of the number of standard model parameters.
\end{tabular}
\vskip0.5cm

There are three discrete symmetries that play a special role, they are\\
$\bullet$ $C$ Charge Conjugation\\
$\bullet$ $P$ Parity\\
$\bullet$ $T$ Time Reversal\\[0.2cm]
QCD and QED conserve $C$, $P$ and $T$ separately. Local Field theory by itself
implies the conservation of $CPT$. 
The fermion and Higgs\footnote{This is only true for the
simplest Higgs sector. In the case of two or more Higgs doublets
a complex phase can appear in the ratios of the vacuum expectation values.
This is known as spontaneous CP-violation or Weinberg's
mechanism.\cite{WeinbergCP}}
part of the SM-Lagrangian conserves $CP$ and $T$
as well.

The only part that violates $CP$ and as a consequence also $T$ is
the Yukawa part.
The Higgs part is responsible for two parameters, the gauge part for three
and the Higgs-Fermion part contains in principle 27 complex parameters,
neglecting Yukawa couplings to neutrinos. In the presence of neutrino masses
and mixings there are more parameters.
Most of the 54 real parameters in the Yukawa sector are unobservable
since they can be removed using field transformations.
After diagonalizing the lepton sector, only the three charged lepton
masses remain. The quark sector can be similarly diagonalized
leading to 6 quark masses, but some parts remain in the difference between
weak interaction eigenstates and mass-eigenstates. The latter is conventionally
put in the couplings of the charged $W$-boson, which is given by
\ba
&\dsp
-\frac{g}{2\sqrt{2}}W^-_\mu 
\left(\overline u^\alpha~\overline c^\alpha~\overline t^\alpha\right)
\gamma^\mu\left(1-\gamma_5\right)
\left(\begin{array}{ccc}V_{ud}&V_{us}&V_{ub}\\
                        V_{cd}&V_{cs}&V_{cb}\\
                        V_{td}&V_{ts}&V_{td}\end{array}\right)
\left(\begin{array}{c}d_\alpha\\s_\alpha\\b_\alpha\end{array}\right)
&\nonumber\\ \nonumber&
\dsp-\frac{g}{2\sqrt{2}}W^-_\mu \sum_{\ell=e,\mu,\tau}
\bar \nu_\ell\gamma^\mu\left(1-\gamma_5\right)\ell
\ea
$C$ and $P$ are broken by the $(1-\gamma_5)$ in the couplings.
$CP$ is broken if $V_{CKM}=\left(V_{ij}\right)$ is (irreducibly) {\em
complex}.

\begin{figure}
\centerline{\includegraphics[width=0.4\textwidth]{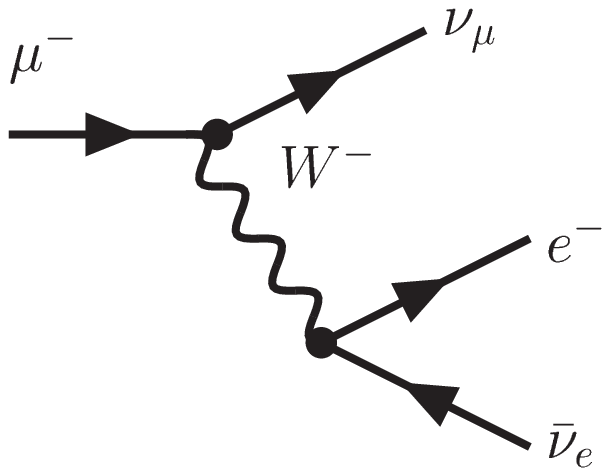}}
\caption{Muon decay: the main source of our knowledge of $g$.}
\end{figure}
The coupling constant $g$ together with the mass $M_W^2$ can be 
determined from the Fermi constant as measured in muon decay.
The relevant corrections are known to two-loop level. The most recent
calculations and earlier references can be found in Ref. \mycite{muon}.
The result is
\be
G_F = \frac{g^2}{8M_W^2} = 1.16639(1)\cdot10^{-5}~\mbox{GeV}^{-2}\,.
\ee

The Cabibbo-Kobayashi-Maskawa matrix
$V_{CKM}=\left(V_{ij}\right)$ results from diagonalizing the quark mass
terms resulting from the Yukawa terms and the Higgs vacuum expectation value.
It is a general unitary matrix but the phases of the quark fields
can be redefined. This allows to remove 5 of the 6 phases present
in a general unitary 3 by 3 matrix.\footnote{We have 6 quark fields
but changing all the anti-quarks by a phase and the quarks by the opposite
phase results in no change in $V_{CKM}$.}
The matrix $V_{CKM}$ thus contains three phases and one mixing angle.
The Particle Data Group preferred parametrization is~\cite{PDG}
\be
\label{CDGparam}
 \left(\begin{array}{ccc}
c_{12}c_{13} & s_{12}c_{13} & s_{13}e^{-i\delta_{13}}\\
-s_{12}c_{23}-c_{12}s_{23}s_{13}e^{i\delta_{13}} &
c_{12}c_{23}-s_{12}s_{23}s_{13}e^{i\delta_{13}} & s_{23}c_{13}\\
s_{12}s_{23}-c_{12}c_{23}s_{13}e^{i\delta_{13}} &
-c_{12}s_{23}-s_{12}c_{23}s_{13}e^{i\delta_{13}} & c_{23}c_{13}\,.
\end{array}\right)
\ee
Using the measured experimental values, see below and~Ref.~\mycite{PDG},
\be
s_{12}=\sin\theta_{12}\approx0.2;\quad
s_{23}\approx0.04\quad\mbox{and}\quad s_{13}\approx0.003
\ee
an approximate parametrization, known as the
Wolfenstein~\cite{Wolfensteinparam} parametrization, can be given.
This is defined via
\be
s_{12}c_{13}\equiv\lambda\,;\quad
s_{23}c_{13}\equiv A\lambda^2\,;\quad\mbox{and}\quad
s_{13}e^{-i\delta_{13}}\equiv A\lambda^3(\rho-i\eta)\,.
\ee
To order $\lambda^4$ the CKM-matrix is
\be
\label{Wolfenstein}
\left(\begin{array}{ccc}
1-\frac{\lambda^2}{2} & \lambda & A\lambda^3(\rho-i\eta)\\
-\lambda & 1-\frac{\lambda^2}{2} & A\lambda^2\\
A\lambda^3(1-\rho-i\eta) & -A\lambda^2 & 1
\end{array}\right)\,.
\ee
$CP$-violation is present if the phase $\delta$ is different from
$0$ or $\pi$ or alternatively $\eta\ne0$.
A way to check for the presence of $CP$ violation
in general set of mass matrices was found by C.~Jarlskog.~\cite{Jarlskog}

A main purpose of this chapter in the book is to describe
how the above picture
of the weak interaction and CKM mixing can be experimentally
tested in the light-quark sector.

\section{Semileptonic decays}
\label{semileptonic}

A first necessary step is the determination of the parameters that occur
in the Yukawa sector of the standard model. Here we concentrate on
the two that can be measured directly in the light quark sector,
$V_{ud}$ and $V_{us}$. As will become very clear, the strong interactions
are always the main theoretical limitation on the precision determination
of these quantities.

\subsection{$V_{ud}$}

The 
underlying idea always the same.
The vector current has matrix-element one at zero momentum transfer
because of the underlying vector Ward identities if all the quark masses
involved in the vector current are equal.

The element $V_{ud}$ is measured in three main sets of decays.
\begin{itemize}
\item{\bf neutron:}
The coupling of the neutron to the weak current is given by
\be
\frac{g_A}{g_V} = -1.2670\pm0.0035\,. 
\ee
where we know that $g_V=1$ at zero momentum transfer.
In addition we use
that the axial current effect is actually measurable via
the angular distribution of the electron. So
$\Gamma(n\to pe^-\bar\nu) \sim G_n^2 (1+3g_A^2/g_V^2) A$\\
with $A$ calculable but containing photon loops.
Using the value for $g_A/g_V$ and the neutron lifetime
of the 2000 particle data book~\cite{PDG}
and the calculations of radiative corrections quoted in Ref.~\mycite{woolcock}
I obtain
\be
|V_{ud}| = 0.9792(40)\,.
\ee
A recent review of the situation for $|V_{ud}|$ is Ref.~\mycite{Vud1}.

At present the errors are dominated by the measurement of $g_A/g_V$.
This could change in the near future.

\item{\bf Nuclear superallowed $\beta$-decays: $0^+\to0^+$}

The main advantages here are that only the vector current can contribute
and that very accurate experimental results are available.
The disadvantage is that the charge symmetry breaking, or isospin, effects
and the photonic radiative corrections are nuclear structure dependent
with an unknown error. The quoted theory errors are such that
the measurements for different nuclei are in contradiction
with each other. In Ref.~\mycite{PDG} they therefore quote
\be
|V_{ud}| = 0.9740(10)\,.
\ee
with a substantial scale factor.

\item{\bf Pion $\beta$-decay: $\pi^+\to\pi^0 e^+\bar\nu$}

The theory here is very clean and improvable using CHPT.\cite{Leutwyler}
The disadvantage is that the
branching ratio is about $10^{-8}$, known to
about 4\%. Experiments at PSI are in progress to get 0.5\%.
\end{itemize}

In the future better experimental precision
in neutron and $\pi$ $\beta$-decay should improve the determination
since they are theoretically under better control.
In any case,
$|V_{ud}|$ is by far the best directly determined CKM-matrix element.
The PDG averages all present measurements to
\be
|V_{ud}| = 0.9735(8)\,.
\ee

\subsection{$V_{us}$}

Again we use the fact that the matrix-element
of a conserved vector current is one at zero momentum. But compared to the
previous subsection we have additional complications:\\[0.2mm]
$\bullet$ The vector current is $\bar s\gamma_\mu u$ so corrections
are $(m_s-m_u)^2$ and not $(m_d-m_u)^2$ so they are naively larger.\\
$\bullet$ A longer extrapolation to the zero-momentum point is
needed.\\[0.2mm]
The relevant semi-leptonic decays can be measured in hyperon or in kaon
decays.

{\bf Hyperon $\beta$-decays (e.g. $\Sigma^-\to n e^-\bar\nu,
\Lambda\to p e^-\bar\nu$)}:
Here there are large theoretical problems.
In CHPT the corrections are large and many new parameters show up. In
addition the series does not converge too well. Curiously enough,
using lowest order CHPT with model-corrections works OK.
For references consult the relevant section of Ref. \mycite{PDG}.
This area needs theoretical work very badly.

{\bf Kaon $\beta$-decays (e.g. $K^+\to \pi^0 e^+\nu$)}:
Both theory and data are old by now. The theory was done by
Leutwyler and Roos \cite{LW}. The analysis uses old-fashioned
photonic loops for the electro-magnetic corrections and one-loop CHPT
for the strong corrections due to quark masses.
Both aspects are at present being improved.~\citetwo{kl3em}{kl3improve}
The latest data are from 1987 and the most precise ones are older.
There is at present a proposal at BNL while KLOE at DAPHNE should also
be able to improve the precision.

The result obtained from the last process is
\be
|V_{us}| = 0.2196\pm0.0023\,.
\ee
We can now use this to test the unitarity relation
\be
|V_{us}|^2+|V_{ud}|^2 = 0.9959\pm0.0019\,.
\ee
$|V_{ub}|^2$ is so small it is not visible in the precision shown here.
The discrepancy is small and should be understood but there is no real cause
for worry at present.

\subsection{Testing CHPT/QCD and determining CHPT parameters}

In the sector of semi-leptonic kaon decays CHPT unfolds all of its
power. An extensive review can be found in Ref.~\mycite{semidaphne}
with a more recent update in Ref.~\mycite{kaon99}. There are of course also
numerous model calculations and other approaches existing.
An example of model calculations using Schwinger-Dyson
equations is in Ref.~\mycite{semimodel}.
Notice that the Schwinger-Dyson equations
themselves do not constitute the model aspect but the assumptions
made in their solutions.

I now simply list the main decays and which quantity they test and/or measure.
\begin{itemize}
\item
$K\to\mu\nu$ ($K_{\ell2}$) : Measurement of $F_K$.
This is known to two loops~\cite{ABT1} in CHPT and the electromagnetic
corrections have also been updated.~\cite{Knecht1}
\item
$K\to\pi\ell\nu$ ($K_{\ell3}$): 
$V_{us}$ and form-factors, see Ref.~\mycite{semidaphne} and references
therein. Work on the two-loop aspects in CHPT and on the
electromagnetic corrections is under way.~\citetwo{kl3improve}{Post}
\item
$K\to\pi\pi\ell\nu$ ($K_{\ell4}$): Form-factors and a main source of
CHPT input parameters. Known at two-loops in CHPT.~\cite{ABT2}
\item
$K\to\pi\ell\nu\gamma$ ($K_{\ell3\gamma}$): Lots of form-factors with
large corrections combining to a final small correction.~\cite{BEG}
\item
$K(\pi)\to\ell\nu\gamma$ ($K(\pi)_{\ell2\gamma}$):
This has two form-factors, one normal and one anomalous. The interference
between them allows to see the sign of the anomaly. This can also be done
in $K_{\ell4}$ and both confirm nicely the
expectations.~\citethree{semidaphne}{BEG}{ABBC}
\end{itemize}

The situation in the semileptonic sector is generally good and we are
talking about precision problems. The medium term limit on the theory
here is the fact that for the quark mass effects we need many rather
high order parameters in CHPT we need to determine somehow. Given the
relative paucity of full higher order calculations at present not much more
can be said. We can expect major progress here in the medium term.

The long term limit will probably come from the fact that the
electromagnetic corrections in the weak decays have free parameters
at higher orders as well. These provide both a virtual photon and 
a virtual 
$W$-boson resulting in an extremely hard problem of matching
long and short distance aspects. Given the difficulty in treating the
same problems with a virtual $W$ only as discussed below, it will probably
take some time before this question is fully settled. 

\section{Non-leptonic Decays: $K\to\pi\pi$ and $K$-$\kob$ mixing}

We have been rather successful in understanding the theory behind the
semi-leptonic decays discussed in the previous section. Basically we always
used CHPT or similar arguments to get at the coupling of the $W$-boson
to hadrons,
problems only occur when we want to go to
very high precision.
A similar simple approach fails completely for non-leptonic
decays. 
I'll first discuss the main qualitative problem that shows up
in trying to estimate these decays, then the phenomenology involved
in mixing phenomena and then proceed in the various steps needed
to actually calculate these processes in the standard model.

\subsection{The $\Delta I=1/2$ rule}

\begin{figure}
\centerline{\includegraphics[width=0.9\textwidth]{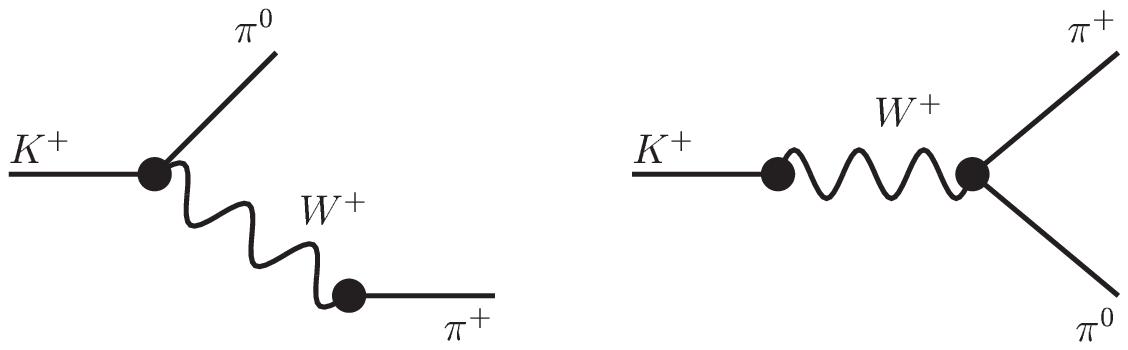}}
\caption{The two naive $W^+$-exchange diagrams for
$K^+\longrightarrow \pi^+\pi^0$.}
\label{figkpipiplus}
\end{figure}

The underlying problem appears when we try to calculate $K\to\pi\pi$
decays in a similar fashion as for the semi-leptonic decays.
For $K^+\to\pi^+\pi^0$ we can draw two Feynman diagrams with a simple $W^+$
exchange as shown in Fig. \ref{figkpipiplus}.
The relevant $W^+$-hadron couplings have all been measured in semi-leptonic
decays and so we have a unique prediction. Comparing this with the
measured decay we get within a factor of two or so.
The approximation described here is known as naive factorization.

\begin{figure}
\centerline{\includegraphics[width=0.5\textwidth]{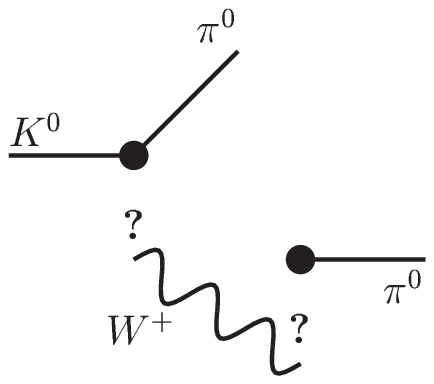}}
\caption{No simple $W^+$-exchange diagram is possible for
$K^0\longrightarrow \pi^0\pi^0$.}
\label{figkpipizero}
\end{figure}

A much worse result appears when we try the same method for the neutral decay
$K^0\to\pi^0\pi^0$. As shown in Fig. \ref{figkpipizero} there is no
possibility to draw diagrams similar to those in Fig. \ref{figkpipiplus}.
The needed vertices always violate charge-conservation.
So we expect that the neutral decay should be small compared
with the ones with charged pions. Well, if we look at the experimental
results we see
\ba
\Gamma(K^0\longrightarrow\pi^0\pi^0)=&
\dsp\frac{1}{2}\Gamma(K_S\longrightarrow\pi^0\pi^0)&=2.3\cdot10^{-12}\mbox{ MeV}
\nonumber\\
\Gamma(K^+\longrightarrow\pi^+\pi^0)=&\dsp
1.1\cdot10^{-14}\mbox{ MeV}&
\ea
So the expected zero one is by far the largest !!!

The same conundrum can be expressed in terms of the
 isospin amplitudes:\,\footnote{Here there are several different sign
and normalization conventions
possible. I present the one used in the work by J.~Prades and myself.}
\ba
A[K^0 \to \pi^0 \pi^0]  &\equiv& \sqrt{\frac{1}{3}} { A_0}
-\sqrt{\frac{2}{3}} \, { A_2} \nonumber\\ 
A[K^0 \to \pi^+ \pi^-]  &\equiv& \sqrt{\frac{1}{3}} { A_0}
+\sqrt{\frac{1}{6}} \, { A_2} \nonumber\\ 
A[K^+ \to \pi^+ \pi^0]  &\equiv& \frac{\sqrt{3}}{2} { A_2}\,.
\ea
The above quoted experimental results can now be rewritten as
\be 
\label{dIhalf1}
\left|\frac{A_0}{A_2}\right|_{\mbox{exp}} = 22.1
\ee
while the naive $W^+$-exchange discussed would give
\be
\label{dInaive}
\left|\frac{A_0}{A_2}\right|_{\mbox{naive}} = \sqrt{2}\,.
\ee
This discrepancy is known as the problem of the $\Delta I=1/2$ rule.
The amplitude which changes the isospin $1/2$ of the kaon to
the zero isospin two pion system is much larger than
the one that changes the isospin to 2 by $3/2$.

Some enhancement is easy to understand from final state $\pi\pi$-rescattering.
Removing these and higher order effects in the light quark masses
one obtains~\cite{KMW}
\be
\label{dIhalf2}
\dsp\left|\frac{A_0}{A_2}\right|_{\chi} = 16.4\,.
\ee
This changes the discrepancy somewhat but is still different by an order
of magnitude from the naive result (\ref{dInaive}). The difference
will have to be explained by pure strong interaction effects
and it is a {\em qualitative} change, not just a quantitative one.

Later we also need the amplitudes with the final state interaction
phase removed via
\be
A_I=-ia_I e^{i\delta_I}
\ee
for $I=0,2$. $\delta_I$ is the angular momentum zero isospin I scattering
phase at the kaon mass.

\subsection{Phenomenology of $K$-$\kob$ mixing and CP-violation}

The $K^0$ and $\kob$\ states are the ones with $\bar sd$ and $\bar ds$
quark content respectively. Up to free phases in these states we can
define the action of $CP$ on these states as
\be
CP |K^0\rangle = -|\kob\rangle\,.
\ee
{}We can construct eigenstates with a definite $CP$ transformation:
\ba
K^0_{1(2)} &=& \frac{1}{\sqrt{2}}
\left(K^0 -(+) \kob\right)\nonumber\\
 CP|K_{1(2)} &=& +(-)|K_{1(2)}\,.
\ea
Now the main decay mode of $K^0$-like states is $\pi\pi$. A two pion state
with charge zero in spin zero is always CP even.
Therefore the decay $K_1\to\pi\pi$ is possible
but $K_2\to\pi\pi$ is {\em impossible}; $K_2\to\pi\pi\pi$ is possible.
Phase-space for the $\pi\pi$ decay is much larger than for the
three-pion final state. Therefore if we start out with a pure $K^0$ or $\kob$\
state, the
$K_2$ component in its wave-function lives much longer than the $K_1$
component.
So after a, by microscopic standards, long time only the $K_2$ component
survives. This was the solution to the two particles with the same
mass and production but very different lifetimes mentioned
in the historical overview, the socalled tau-theta puzzle.

In the early sixties, as you see it pays off to do precise experiments,
one actually measured \cite{CCFT}
\be
\frac{\Gamma(K_L\to\pi^+\pi^-)}{\Gamma(K_L\to\mbox{all})}=
(2\pm0.4)\cdot10^{-3}\,.
\ee
So we see that {\em $CP$ is violated}\,.

This leaves us with the questions:
\begin{itemize}
\item Does $K_1$ turn in to $K_2$ ($CP$-violation in mixing or
indirect $CP$ violation)?
\item Does $K_2$ decay directly into $\pi\pi$ (direct $CP$ violation)?
\end{itemize}
In fact, the 
answer to both is {\em YES}\, and is major qualitative test
of the standard model Higgs-fermion sector and the entire $CKM$-picture
of $CP$-violation.

Let us now describe the $K^0\kob$ system in somewhat more detail.
The Hamiltonian, seen  as a two state system, is given by
\be
i\frac{d}{dt}
\left(\begin{array}{c}K^0\\[1mm]\kob\end{array}\right)
=
\left(\begin{array}{cc}M_{11}-\frac{i}{2}\Gamma_{11}&
M_{12}-\frac{i}{2}\Gamma_{12}\\[1mm]
M_{21}-\frac{i}{2}\Gamma_{21}&M_{22}-\frac{i}{2}\Gamma_{22}\end{array}\right)
\left(\begin{array}{c}K^0\\[1mm]\kob\end{array}\right)
\ee
where $M = \left(M_{ij}\right)$ and $\Gamma =\left(\Gamma_{ij}\right)$
are hermitian two by two matrices. The Hamiltonian itself is allowed to have a
non-hermitian part since we do not conserve probability here. The kaons
themselves can decay and the anti-hermitian part $\Gamma$ describes the
decays of the kaons to the ``rest of the universe.''

CPT implies, a derivation can be found in Ref.~\mycite{eduardo},
\ba
M_{11} = M_{22}\quad&& \Gamma_{11}=\Gamma_{22}\nonumber\\
M_{12} = M_{21}^*\quad&&
\Gamma_{12}=\Gamma_{21}^*
\ea
and this assumption can in fact be relaxed for tests of CPT.

Diagonalizing the Hamiltonian we obtain
\be
K_{S(L)} = \frac{1}{\sqrt{1+|\tilde\varepsilon|^2}}
\left(K_{1(2)}+\tilde\varepsilon K_{2(1)}\right)
\ee
as physical propagating states. Notice that they are not orthogonal,
which is allowed since the Hamiltonian is not hermitian.

We define the following observables
\ba
\eta_{+-} &\equiv& \frac{A(K_L\to\pi^+\pi^-)}{A(K_S\to\pi^+\pi^-)}
\,,\nonumber\\
\eta_{00} &\equiv& \frac{A(K_L\to\pi^0\pi^0)}{A(K_S\to\pi^0\pi^0)}
\,,\nonumber\\
\varepsilon &\equiv& \frac{A(K_L\to(\pi\pi)_{I=0})}{A(K_S\to(\pi\pi)_{I=0})}
\ea
and
\be
\varepsilon^\prime = \frac{1}{\sqrt{2}}\left(
\frac{A(K_L\to(\pi\pi)_{I=2})}{A(K_S\to(\pi\pi)_{I=0})}-
\varepsilon
\frac{A(K_S\to(\pi\pi)_{I=2})}{A(K_S\to(\pi\pi)_{I=0})}\right)\,.
\ee
The latter has been specifically constructed to remove the 
$K^0$-$\kob$\ transition.
$|\varepsilon|$, $|\eta_{+-}|$ and
$|\eta_{00}|$ are directly measurable as ratios of decay rates.

We now make a series of approximations that are experimentally
valid,
\ba
 |\im a_0|,|\im a_2| &<<& |\re a_2| << |\re a_0| \nonumber\\
 |\varepsilon|,|\tilde \varepsilon| &<<& 1\nonumber\\
 \quad |\varepsilon^\prime| &<<& | \varepsilon|\,,
\ea
to obtain the usually quoted expressions
\be
\varepsilon^\prime = \frac{i}{\sqrt{2}}e^{i(\delta_2-\delta_0)}
\frac{\re a_2}{\re a_0}\left(\frac{\im a_2}{\re a_2}-\frac{\im a_0}{\re a_0}
\right)
\ee
and
\be
\varepsilon = \tilde\varepsilon+ i\frac{\im a_0}{\re a_0}\,.
\ee
For the latter we
use $\Delta m = m_L-m_S \approx \frac{\Delta\Gamma}{2}$ and 
$\Gamma_L << \Gamma_S$ and the fact that
$\Gamma_{12}$ is dominated by $\pi\pi$ states and get
\be
\varepsilon = \frac{1}{\sqrt{2}}e^{i\pi/4}
\left(\frac{\im M_{12}}{\Delta m}+\frac{\im a_0}{\re a_0}\right)\,.
\ee
Putting all the above together we finally get to
\be
\eta_{+-} = \varepsilon+\varepsilon^\prime
\quad\mbox{and}\quad
\eta_{00} = \varepsilon-2\varepsilon^\prime\nonumber
\ee
Where you can see that $\varepsilon$ describes the indirect part
and $\varepsilon^\prime$ the direct part, since the mixing contribution would
be the same for $\eta_{+-}$ and $\eta_{00}$.

The real part of $\tilde\varepsilon$ can be measured in semi-leptonic
decays via the ratio
\be
\delta = 
\frac
{\dsp \Gamma(K_L\to\pi^-\ell^+\nu_\ell)-\Gamma(K_L\to\pi^+\ell^-\bar\nu_\ell)}
{\dsp \Gamma(K_L\to\pi^-\ell^+\nu_\ell)+\Gamma(K_L\to\pi^+\ell^-\bar\nu_\ell)}
\,.
\ee
The latter relation assumes the $\Delta S=\Delta Q$ rule which is satisfied
to a very good precision in the Standard Model. The sign of the lepton
charge thus tells us whether $K_L$ decayed as a $K^0$ or a $\kob$.

\subsection{Experimental results}

Experimentally,~\citetwo{PDG}{Kessler}
\be
2\,\re\varepsilon\approx\delta = 0.00331\pm0.00006\,.
\ee
including the recent preliminary KTeV result.

$\varepsilon$ is well known.~\cite{PDG}
\be
|\varepsilon| = (2.271\pm0.017)\cdot 10^{-3}\,,
\ee
as determined by several high precision experiments which are in good
agreement with each other.

The experimental situation on $\epspeps$ was unclear for a long time.
Two large experiments, NA31 at CERN and E731 at FNAL,
obtained conflicting results
in the mid 1980's. Both groups have since gone on and build improved versions
of their detectors, NA48 at CERN and KTeV at FNAL.
$\epspeps$ is
measured via the double ratio
\ba
\re\left(\frac{\varepsilon^\prime}{\varepsilon}\right)
&=& \frac{1}{6}\left\{
1-\left|\frac{\eta_{00}}{\eta_{+-}}\right|^2\right\}
\nonumber\\
&=& \frac{1}{6}\left\{
1-\frac{\dsp \Gamma(K_L\to\pi^+\pi^-)/\Gamma(K_S\to\pi^+\pi^-)}
{\dsp \Gamma(K_L\to\pi^0\pi^0)/\Gamma(K_S\to\pi^0\pi^0)}
\right\}\,.
\ea
The two main experiments follow a somewhat different strategy in
measuring this double ratio, mainly in the way the relative normalization
of $K_L$ and $K_S$ components is treated.
After some initial disagreement with the first results, KTeV has
reanalyzed their systematic errors and the
situation for $\epspeps$ is now quite clear.
We show the recent results in Table \ref{tabepspeps}.
The data are taken from Ref.~\mycite{epspeps} and the recent
reviews in the Lepton-Photon conference.~\citetwo{Kessler}{NA4801} 
\begin{table}
\begin{center}
{\begin{tabular}{|cc|}
\hline
NA31 & $(23.0\pm6.5)\times 10^{-4}$\\
E731 & $(7.4\pm5.9)\times 10^{-4}$\\
\hline
KTeV 96 & $(23.2\pm4.4)\times10^{-4}$\\
KTeV 97 & $(19.8\pm2.9)\times10^{-4}$\\
NA48 97 & $(18.5\pm7.3)\times10^{-4}$\\
NA48 98+99 & $(15.0\pm2.7)\times10^{-4}$\\
\hline
ALL & $(17.2\pm1.8)\times10^{-4}$\\
\hline
\end{tabular}}
\end{center}
\caption{Recent results on $\varepsilon^\prime/\varepsilon$. The years
refer to the data sets.}
\label{tabepspeps}
\end{table}

\subsection{Standard Model Diagrams}

\begin{figure}
\centerline{\includegraphics[width=0.7\textwidth]{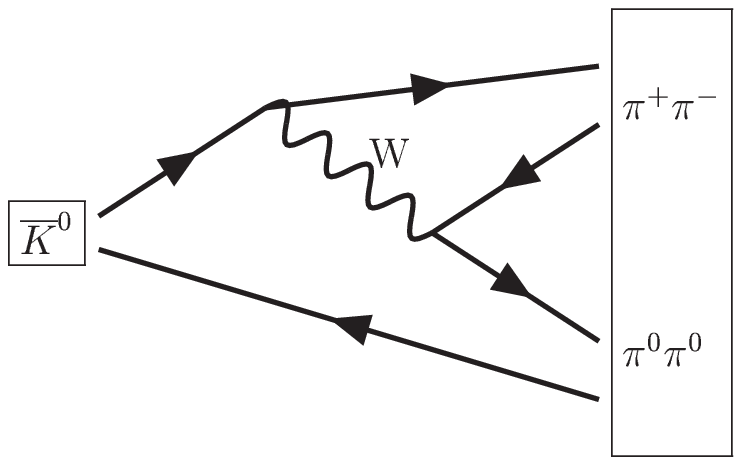}}
\caption{\label{figW}
The $W$-exchange diagram, sometimes referred to as the spectator mechanism.
Extra gluons etc. are not shown.}
\end{figure}

\begin{figure}
\centerline{\includegraphics[width=0.7\textwidth]{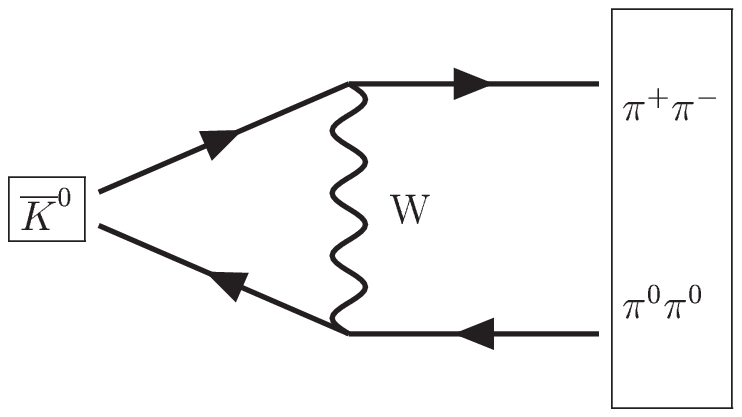}}
\caption{\label{figW2}
The $W$-exchange diagram, sometimes referred to as the annihilation mechanism.
Extra gluons etc. are not shown.}
\end{figure}

The weak nonleptonic decays all happen via the exchange of a $W$-boson.
The main diagram is depicted in Fig.~\ref{figW} where we have only
indicated one possible routing of the quark legs and no extra gluons.
This figure should be understood as an indication of the type of diagrams
responsible. The contribution depicted here is sometimes referred to
as the spectator contribution, but that name normally implies some more
approximations. The other simple $W$-exchange diagram is similarly known
as the annihilation contribution and is depicted in Fig.~\ref{figW2}.
These diagrams, and the ones with extra gluons and light quarks, are the
ones responsible for the main part of the decay rate.

\begin{figure}
\centerline{\includegraphics[width=0.6\textwidth]{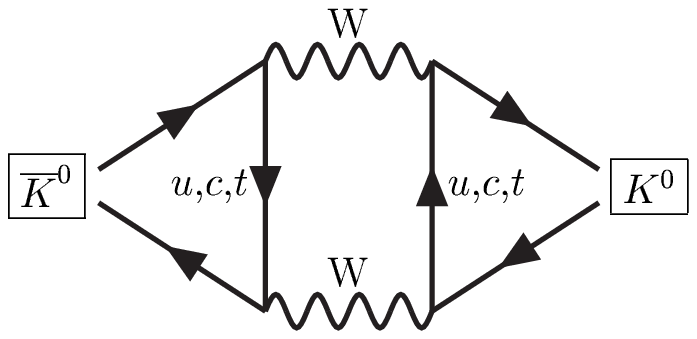}}
\caption{\label{figbox}
 The box diagram contribution to $K^0\kob$\ mixing.
Crossed versions and diagrams with extra gluons etc. are not shown.}
\end{figure}

$CP$-violation with the CKM mechanism goes via another class of diagrams.
The complex phase in the CKM matrix can be removed as soon as effectively
not all three of the generations contribute. The diagrams thus need
to have a component in it that involves all three generations in order to
contribute to $CP$-violation.
The set of diagrams, depicted schematically in Fig.~\ref{figbox},
responsible for $K^0\kob$\ mixing are known as box diagrams.
It is the presence of the virtual intermediate quark lines
of up, charm and top quarks that produces the $CP$-violation. 
\begin{figure}
\centerline{\includegraphics[width=0.9\textwidth]{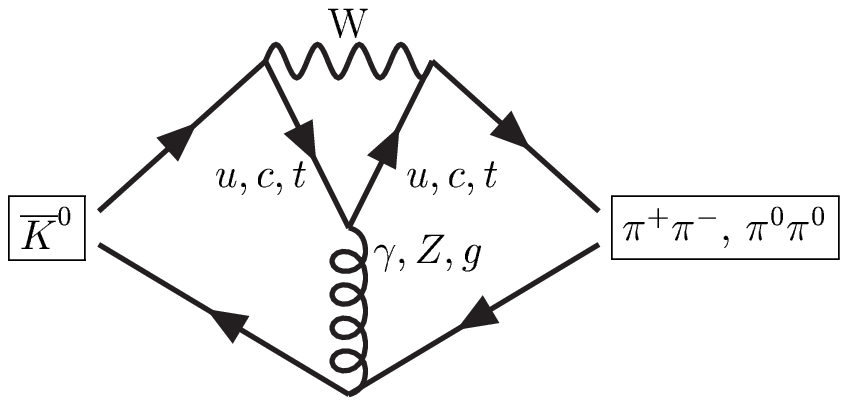}}
\caption{\label{figpenguin} The Penguin diagram contribution to $K\to\pi\pi$.
Extra gluons and crossed versions etc. are not shown.}
\end{figure}

The Penguin diagram shown in Fig. \ref{figpenguin} contributes to the direct
$CP$-violation as given by $\varepsilon^\prime$.
Again, $W$-couplings to all three generations show up so CP-violation
is possible in $K\to\pi\pi$. This is a qualitative prediction of the
standard model and borne out by experiment.
The main problem is now to embed these diagrams and the simple $W$-exchange
in the full strong interaction. The $\Delta I=1/2$ rule shows that there
will have to be large corrections to the naive picture.

\subsection{The first QCD correction}
\label{firstQCD}

\begin{figure}
\centerline{\includegraphics[width=0.5\textwidth]{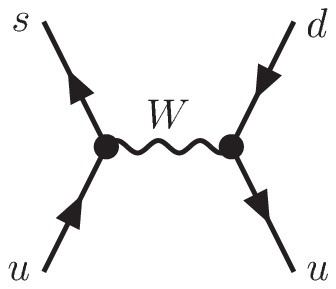}}
\caption{The $W$-exchange depicted at the quark level.\label{figW3}}
\end{figure}

\begin{figure}
\centerline{\includegraphics[width=0.5\textwidth]{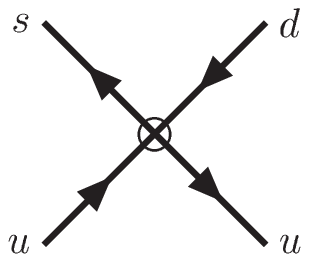}}
\caption{The effect of the local operators
depicted at the quark level.\label{figQ1}}
\end{figure}

\begin{figure}
\centerline{\includegraphics[width=0.5\textwidth]{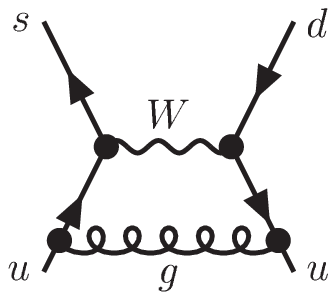}}
\caption{A diagram with a gluonic correction to tree level $W$-exchange.
\label{figWglue}}
\end{figure}

The naive factorization approach we described earlier can reformulated
in a different fashion which makes it easier to generalize later on.
Since the kaon mass is so much lower than the $W$ mass, we expect that
at first approximation we can neglect the momentum dependence of the
$W$-propagator. So we replace the effect of $W$ exchange of
Figs.~\ref{figW} and \ref{figW2} by the effect of a local four-quark operator.
So we replace at the quark level Fig.~\ref{figW3} by
the operators depicted in Fig.~\ref{figQ1}.
The $W$ exchange at tree level can be described by the effective
Hamiltonian
\be
 {\cal H}_{\mbox{eff}}^{\mbox{tree}} = \sum_i C_i(\mu) Q_i(\mu)
\label{defHefftree}
\dsp{\cal H}_{\mbox{eff}} = \frac{G_F}{\sqrt{2}} V_{ud}V_{us}^*
\sum_i z_i Q_i^u\,.
\ee
The coefficients $z_i$ are real and
 the CKM-matrix elements occurring are shown explicitly.
The four-quark operators $Q_i^u$ are defined by
\ba
\label{defQiu}
Q_1^u &=& 
(\bar s_\alpha\gamma_\mu u_\beta)_L (\bar u_\beta\gamma^\mu d_\alpha)_L\,,
\nonumber\\
Q_2^u &=&
 (\bar s_\alpha\gamma_\mu u_\alpha)_L (\bar u_\beta\gamma^\mu d_\beta)_L\,,
\ea
with $(\bar q \gamma_\mu q^\prime)_L = \bar q \gamma_\mu(1-\gamma_5)q^\prime$.
In order to have the same matrix elements as
$W$-exchange at tree level up to terms of order $m_K^2/m_W^2$ we need to set
\be
z_1 = 0\quad\mbox{and}\quad z_2 = 1\,.
\ee
The effect of gluonic interactions can now be taken into account.
Diagrams like the one depicted in Fig.~\ref{figWglue} contribute as well.
The total results is finite, after the renormalization of $\alpha_S$
is duly taken into account. The result, again up to terms
of order $m_K^2/m_W^2$, can be taken into account by replacing $z_1$
and $z_2$ by
\ba
\label{zioneloop}
z_1 &=& -\frac{\alpha_S}{4\pi} 3 \ln\left(\frac{m_W^2}{\mu^2}\right)
\approx 0.84\,,
\nonumber\\
z_2 &=& 1+\frac{\alpha_S}{4\pi} \ln\left(\frac{m_W^2}{\mu^2}\right)
\approx 1.28\,.
\ea
For the numerical results we have used $\mu\approx 1~$GeV and
$\alpha_S\approx 0.4$.

What conclusions can we draw from this? First of all, the corrections are
rather large because of the presence of the large logarithmic term. This
we will have to take care of more accurately as described below.
A second consequence is that we have already gone some way towards
solving the $\Delta I=1/2$ rule. The operators
\be
Q_{\pm} = \frac{1}{2}\left(Q_2^u\pm Q_1^u\right)
\ee
have a different isospin behaviour. $Q_-$ is pure isospin $1/2$ while
$Q_+$ has both isospin $1/2$ and $3/2$ components.
The tree level result gave both of them with coefficients $z_\pm = 1$
while Eq. (\ref{zioneloop}) gives
\be
z_- \approx 2.1 \quad\mbox{and}\quad z_+ \approx 0.44\,.
\ee
We see a significant enhancement of the $\Delta I=1/2$ component and a
suppression of the $\Delta I=3/2$ component.
We have not said how one gets from the quark level to the meson level here.
This will be discussed in more detail in Sect. \ref{longdistance}.

\subsection{The steps from quarks to mesons in the weak interaction}

We need to extend the calculation of Sect.~\ref{firstQCD}
in several ways. We only took into account the simplest gluonic diagrams,
we should preferably go to higher orders and the large logarithms need
to be treated to all orders if possible.
The latter can be done since the large logarithms contain the subtraction scale
$\mu$. The dependence on $\mu$ of an observable must vanish, which means
that coefficients of the type $z_i$ must depend on $\mu$ in such a way that
when a matrix element between physical states is taken the $\mu$-dependence
cancels. This can be used to calculate the variation of the $z_i$ with $\mu$
in a way which includes all term of order $\left(\alpha_S\ln\mu^2\right)^n$
with a one-loop calculation and the techniques of the renormalization group.
It can also be improved by going to higher orders.
We proceed thus in several steps, we replace the exchange of $W$'s by
a series of operators at a scale $\mu$ such that the logarithms
of the type $\ln(m_W^2/\mu^2)$ are small. The coefficients $z_i$ (or
more general $C_i$) are universal and can be calculated by comparing matrix
elements of $W$-exchange with the matrix elements of the local operators.
This correspondence is {\em independent} of the choice of matrix elements,
so the external states can be chosen to minimize the effort needed for the
calculation.

The $\mu$ dependence of the $C_i$ is then exploited via the
renormalization group to resum all large logarithms to the order desired,
or rather to the order the calculations can be practically performed.
The last step is then to take the matrix elements of the operators
at a low scale $\mu$ which should avoid large logarithms of the type
$\ln(m_K^2/\mu^2)$. 

The three steps of the
full calculation are depicted in 
Fig. \ref{figsteps}.
\begin{figure}
\begin{tabular}{ccc}
ENERGY SCALE & FIELDS & Effective Theory\\
\hline
\\[2mm]
$M_W$ & \framebox{\parbox{5cm}{\begin{center}
$W,Z,\gamma,g$;\\$\tau,\mu,e,\nu_\ell$;\\
$t,b,c,s,u,d$\end{center}
}} & Standard Model\\[2mm]
 & {\em $\Downarrow$ using OPE} & \\[2mm]
$\lesssim m_c$ & \framebox{\parbox{5cm}{\begin{center}
$\gamma,g$; $\mu,e,\nu_\ell$;\\ $s,d,u$\end{center}
}} & \parbox{2cm}{QCD,QED,\\${\cal H}_{\mbox{eff}}^{|\Delta S|=1,2}$}\\[2mm]
  & {\em$\Downarrow$ ???} & \\[2mm]
$M_K$  & \framebox{\parbox{5cm}{\begin{center}$\gamma$; $\mu,e,\nu_\ell$;\\
$\pi$, $K$, $\eta$\end{center}}} & CHPT\\ \\\hline
\end{tabular}
\caption{A schematic exposition of the various steps in the calculation
of nonleptonic matrix-elements.}
\label{figsteps}
\end{figure}
First we integrated out the heaviest particles step by step using Operator
Product Expansion methods. The steps OPE we describe in the next subsections
while step {\em ???} we will split up in more subparts later.

\subsection{Step I: from SM to OPE}

\begin{figure}
\centerline{\includegraphics[width=0.7\textwidth]{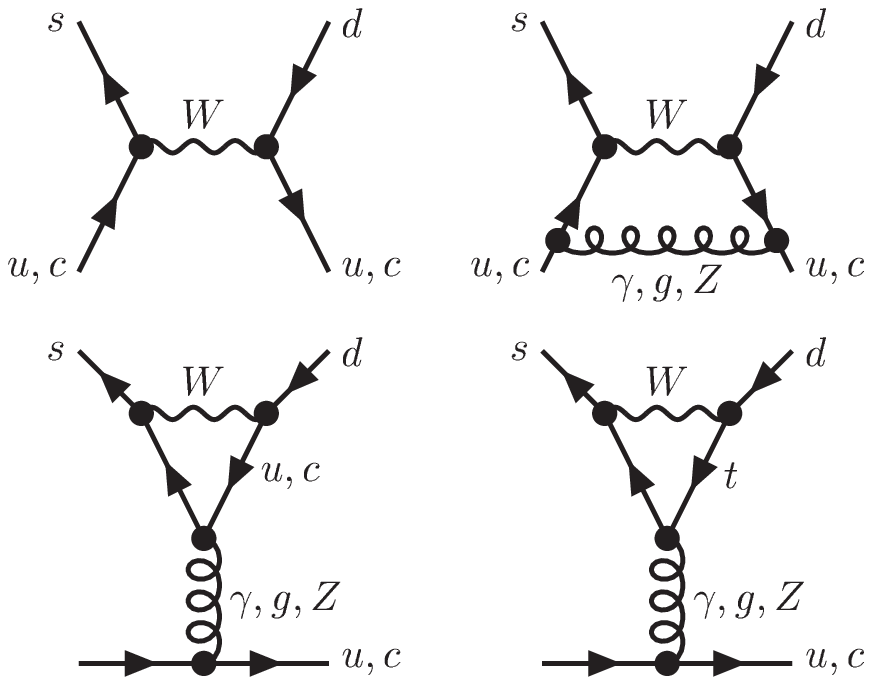}}
\caption{The standard model diagrams to be calculated at a high scale.
\label{figSM}}
\end{figure}

The first step for the processes we discuss in this review
concerns the
 standard model diagrams of Fig. \ref{figSM}.
We replace their effect
with a contribution of an effective Hamiltonian given by
\ba
\label{defHeff}
 {\cal H}_{\mbox{eff}} &=& \sum_i C_i(\mu) Q_i(\mu)
\nonumber\\
&=& \frac{G_F}{\sqrt{2}} V_{ud}V_{us}^*
\sum_i\left(z_i-y_i\frac{V_{td}V_{ts}^*}{V_{ud}V_{us}^*}\right)Q_i\,.
\ea
In the last part we have real coefficients $z_i$ and $y_i$ and
 the CKM-matrix elements occurring are shown explicitly.
The four-quark operators $Q_i$ are defined by
\ba
\label{defQi}
Q_{1,2} &=& Q_{1,2}^u-Q_{1,2}^c
\,,\nonumber\\
Q_1^u &=& 
(\bar s_\alpha\gamma_\mu u_\beta)_L (\bar u_\beta\gamma^\mu d_\alpha)_L
\,,\nonumber\\
Q_1^c &=&
 (\bar s_\alpha\gamma_\mu c_\beta)_L (\bar c_\beta\gamma^\mu d_\alpha)_L
\,,\nonumber\\
Q_2^u &=&
 (\bar s_\alpha\gamma_\mu u_\alpha)_L (\bar u_\beta\gamma^\mu d_\beta)_L
\,,\nonumber\\
Q_2^c &=&
 (\bar s_\alpha\gamma_\mu c_\alpha)_L (\bar c_\beta\gamma^\mu d_\beta)_L
\,,\nonumber\\
Q_3 &=& (\bar s_\alpha\gamma_\mu d_\alpha)_L
\sum_{q=u,d,s,c,b} (\bar q_\beta\gamma^\mu q_\beta)_L
\,,\nonumber\\
Q_4 &=& (\bar s_\alpha\gamma_\mu d_\beta)_L
\sum_{q=u,d,s,c,b} (\bar q_\beta\gamma^\mu q_\alpha)_L
\,,\nonumber\\
Q_5 &=& (\bar s_\alpha\gamma_\mu d_\alpha)_L
\sum_{q=u,d,s,c,b} (\bar q_\beta\gamma^\mu q_\beta)_R
\,,\nonumber\\
Q_6 &=& (\bar s_\alpha\gamma_\mu d_\beta)_L
\sum_{q=u,d,s,c,b} (\bar q_\beta\gamma^\mu q_\alpha)_R
\,,\nonumber\\
Q_7 &=& (\bar s_\alpha\gamma_\mu d_\alpha)_L
\sum_{q=u,d,s,c,b} \frac{3}{2}e_q(\bar q_\beta\gamma^\mu q_\beta)_R
\,,\nonumber\\
Q_8 &=& (\bar s_\alpha\gamma_\mu d_\beta)_L
\sum_{q=u,d,s,c,b} \frac{3}{2}e_q(\bar q_\beta\gamma^\mu q_\alpha)_R
\,\nonumber\\
Q_9 &=& (\bar s_\alpha\gamma_\mu d_\alpha)_L
\sum_{q=u,d,s,c,b} \frac{3}{2}e_q(\bar q_\beta\gamma^\mu q_\beta)_L
\,,\nonumber\\
Q_{10} &=& (\bar s_\alpha\gamma_\mu d_\beta)_L
\sum_{q=u,d,s,c,b} \frac{3}{2}e_q(\bar q_\beta\gamma^\mu q_\alpha)_L
\ea
with $(\bar q\gamma_\mu q^\prime)_{(L,R)} 
= \bar q \gamma_\mu(1\mp\gamma_5)q^\prime$;
$\alpha$ and $\beta$ are colour indices.

\begin{figure}
\centerline{\includegraphics[width=0.95\textwidth]{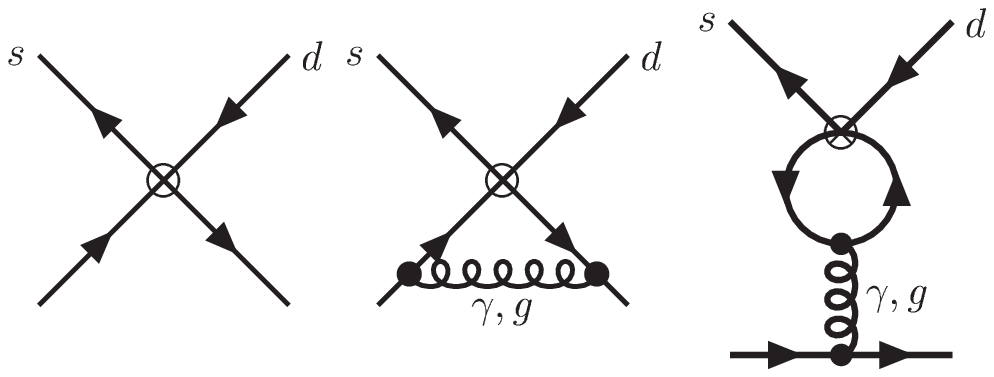}}
\caption{The diagrams needed for the
matrix-elements calculated at a scale $\mu_H\approx m_W$
using the effective Hamiltonian.}
\label{figOPE}
\end{figure}
We calculate now matrix elements between quarks and gluons in the
standard model using the diagrams of Fig. \ref{figSM}
and equate those to the same matrix-elements calculated using the
effective Hamiltonian of Eq. (\ref{defHeff}) and the diagrams
of Fig. \ref{figOPE}. This determines the value of the $z_i$ and $y_i$.
The top quark and the $W$ and $Z$ bosons are integrated out all at the
same time. There should be no large logarithms present due to that.
The scale $\mu = \mu_H$ in the diagrams of Fig.~\ref{figOPE}
of the OPE expansion diagrams
should be chosen of the order of the $W$ mass.
The scale $\mu_W$ in the Standard Model diagrams of Fig.~\ref{figSM}
should be chosen of the same order.

{\underline{Notes}:}\\
$\bullet$ In the Penguin diagrams
$CP$-violation shows up since all 3 generations
are present.\\
$\bullet$ The equivalence is done by calculating matrix-elements between
{\em Quarks and Gluons}\\
$\bullet$ The SM part is $\mu_W$-independent to $\alpha_S^2(\mu_W)$.\\
$\bullet$ OPE part: The $\mu_H$ dependence of
$C_i(\mu_H)$ cancels the $\mu_H$ dependence of the diagrams
to order $\alpha_S^2(\mu_H)$.

This procedure gives at $\mu_W = \mu_H=M_W$
in the NDR-scheme~\footnote{The precise definition
of the four-quark operators $Q_i$ comes in here as well. See the lectures by
Buras \cite{Buras1} for a more extensive description of that.}
the numerical values given in Table \ref{tabzimw}.
\begin{table}
\begin{center}
\begin{tabular}{|ccc|}
\hline
$z_1$   & 0.053     &  { $g,\gamma$-box} \\
$z_2$	& 0.981	    &{ $W^+$-exchange  $g,\gamma$-box} \\
\hline
$y_3$	& 0.0014    & {$g,Z$-Penguin  $WW$-box} \\
$y_4$	&$-$0.0019  & {$g$-Penguin}\\
$y_5$	& 0.0006    & {$g$-Penguin}\\
$y_6$	&$-$0.0019  & {$g$-Penguin}\\
$y_7$	& 0.0009    & {$\gamma,Z$-Penguin}\\
$y_8$	& 0.	    & \\
$y_9$	& $-$0.0074 & {$\gamma,Z$-Penguin  $WW$-box}\\
$y_{10}$& 0.        & \\
\hline
\end{tabular}
\end{center}
\caption{The Wilson coefficients and their main source at
the scale $\mu_H=m_W$ in the NDR-scheme.}
\label{tabzimw}
\end{table}
In the same table I have
given the main source of these numbers. Pure tree-level $W$-exchange
would have only given $z_2=1$ and all others zero.
Note that the coefficients from $\gamma,Z$ exchange are similar to the gluon
exchange ones since $\alpha_S$ at this scale is not very big.

\subsection{Step II}
\label{shortdistance}

Now comes the main advantage of the OPE formalism. Using the
renormalization group equations we can  calculate
the change with $\mu$ of the $C_i$ thus resumming the
$\log\left(m_W^2/\mu^2\right)$ effects.

The renormalization group equation (RGE) for the
strong coupling is
\be
\mu\frac{d}{d\mu}g_S(\mu) = \beta(g_S(\mu))
\ee
and for the Wilson coefficients
\be
\label{RGE}
\mu\frac{d}{d\mu}C_i(\mu) = \gamma_{ji}(g_S(\mu),\alpha) C_j(\mu)\,.
\ee
$\beta$ is the QCD beta function for the running coupling.

The coefficients
$\gamma_{ij}$ are the elements
of the anomalous dimension matrix $\hat\gamma$. This can be derived from
the infinite parts of loop diagrams and this has been done to
one \cite{one-loop} and two loops.\cite{two-loop}
The series in $\alpha$ and $\alpha_S$ is known to
\be
\hat\gamma = \hat\gamma^0_S \frac{\alpha_S}{4\pi} 
            + \hat\gamma^1_S \left(\frac{\alpha_S}{4\pi}\right)^2
            + \hat\gamma_e \frac{\alpha}{4\pi}
            + \hat\gamma_{se}\frac{\alpha_S}{4\pi} \frac{\alpha}{4\pi}
            +\cdots 
\ee
Many subtleties are involved in this calculation.\citetwo{Buras1}{two-loop}
They all are related to the fact that everything at higher loop orders need
to be specified correctly, and many things which are equal at tree
level are no longer so in $d\ne4$ and at higher loops.
The main subtleties are:\\
$\bullet$ 
{The definition of $\gamma_5$ is important: 
Naive dimensional regularization (NDR), $\gamma_5$
anticommutes with  $\gamma_\mu$, versus 't~Hooft-Veltman}
(HV), $\gamma_5$ anticommutes with the 4-dimensional part of $\gamma_\mu$
but commutes with the $d-4$-dimensional part.\\
$\bullet$
{Fierzing is important: special care needs to
be taken how to write the operators}\\
$\bullet$
{Evanescent operators}\\
$\bullet$ The definition of the axial current. The NDR one is
conserved for massless quarks but the HV one has an ambiguity since
$\{\gamma_\mu,\gamma_5\}\ne0$.

An introductory review to this is Ref.~\mycite{Buras1} and
a review with numerical results for all the Wilson coefficients
is Ref.~\mycite{twoloopreview}. The analytical solution of the equations
(\ref{RGE}) to two-loop order is somewhat tricky as described
in the quoted references. The numbers below are obtained by numerically
integrating Eq. (\ref{RGE}).

We need to perform the following steps to get down to a scale $\mu_{OPE}$
somewhere around 1~GeV. Starting from the values of $z_i$ and $y_i$ given
in Table~\ref{tabzimw} at the scale {\boldmath$\mu_H$}.
\begin{enumerate}
\item
Solve Eqs. (\ref{RGE}) numerically and/or analytically; run from 
{\boldmath${\mu_H}$}
to {\boldmath${\mu_b}$}
\item
At {\boldmath$\mu_b$}$\approx m_b$ remove $b$-quark and do matching to
theory without $b$. This is done by calculating matrix elements of the
effective Hamiltonian in the five quark picture and in the four quark picture
and putting them equal. This does lead to discontinuities in the values of the
$C_i$.
\item
run down from {\boldmath$\mu_b$} to {\boldmath$\mu_c$}$\approx m_c$
\item
At {\boldmath$\mu_c$} remove the $c$-quark and do matching to
the theory without $c$.
\item
run from {\boldmath$\mu_c$} to {\boldmath$\mu_{\mbox{OPE}}$}
\end{enumerate}
This way we summed {\em all} large logarithms including $m_W$, $m_Z$, $m_t$,
$m_b$ and $m_c$. This is easily done this way, impossible
otherwise.

Notice that we had lots of scales {\boldmath$\mu_i$}.
In principle nothing depends on any of them and varying them gives an
indication of the neglected higher order corrections.

With the inputs
$m_t(m_t)=166~GeV$, $\alpha=1/137.0$, $\alpha_S(m_Z) = 0.1186$
which led to the initial conditions shown in table \ref{tabzimw}, we can
perform the above procedure down to $\mu_{OPE}$.
Results for 900~MeV are shown in columns two and three of Table \ref{tabzimu}.
\begin{table}
\begin{center}
\begin{tabular}{|c|cc|cc|}
\hline
i        & $z_i$ & $y_i$ & $z_i$ & $y_i$ \\
         &\multicolumn{2}{c}{$\mu_{OPE} = 0.9$ GeV}\vline
&\multicolumn{2}{c}{$\mu_X =0.9$ GeV}\vline\\
\hline
$z_{1}$ &$-$0.490           & 0.                & $-$0.788        & 0.\\
$z_{2}$	&1.266 		    & 0.                & 1.457           & 0. \\
$z_{3}$	&0.0092		    &  0.0287           & 0.0086          & 0.0399\\
$z_{4}$	&$-$0.0265	    & $-$0.0532         & $-$0.0101       & $-$0.0572\\
$z_{5}$	& 0.0065	    & 0.0018            & 0.0029          & 0.0112\\
$z_{6}$	&$-$0.0270	    & $-$0.0995         &  $-$0.0149      & $-$0.1223\\
$z_{7}$	& 2.6$~10^{-5}$	    &$-$0.9$~10^{-5}$   & 0.0002          &$-$0.00016\\
$z_{8}$	& 5.3$~10^{-5}$	    & 0.0013            &  6.8$~10^{-5}$  & 0.0018\\
$z_{9}$	& 5.3$~10^{-5}$	    & $-$0.0105         &  0.0003         & $-$0.0121\\
$z_{10}$&$-$3.6$~10^{-5}$   &0.0041             & $-$8.7$~10^{-5}$& 0.0065\\
\hline
\end{tabular}
\end{center}
\caption{The Wilson coefficients
$z_i$ and $y_i$ at a scale $\mu_{\mbox{OPE}}=$ 900~MeV
in the NDR scheme and in the $X$-boson scheme at $\mu_X =$ 900~MeV.}
\label{tabzimu}
\end{table}
Notice that $z_1$ and $z_2$ have changed much from $0$ and $1$ and are also
significantly different from the simple one-loop estimate.
This is the short-distance contribution to the $\Delta I=1/2$ rule.
We also see a large enhancement of $y_6$ and $y_8$, which will
lead to our value of $\varepsilon^\prime$.

A similar exercise can be performed for $K^0$-$\kob$\ mixing.\cite{Herrlich}
This yields the effective Hamiltonian
\be
{\cal H}_{\mbox{eff}}^{\Delta S=2} = C_{\Delta S=2}
\left(\bar s_\alpha\gamma_\mu d_\alpha\right)_L
\left(\bar s_\beta\gamma_\mu d_\beta\right)_L
\ee
with
\ba
C_{\Delta S=2} &=& \frac{G_F^2 M_W^2}{16\pi^2}
\left[\lambda_c^2\eta_1 S_0(x_c)
+\lambda_t^2\eta_2 S_0(x_t)+2\lambda_c\lambda_t S_0(x_c,x_t)\right]
\nonumber\\
&&\times
\alpha_S^{(-2/9)}(\mu)
\left(1+\frac{\alpha_S(\mu)}{4\pi} J_3\right)
\ea
and
\be
x_c =\frac{m_c}{M_W^2}\quad
x_t =\frac{m_t}{M_W^2}\quad 
\lambda_i=-V_{id}V_{is}^*\quad
J_3(n_f=3) = \frac{307}{162}\,.
\ee
The functions were first derived by Inami and Lin~\cite{InamiLim}
\ba
S_0(x_c)&\approx&x_c\,,
\nonumber\\
S_0(x_t)&=&\frac{4 x_t-11 x_t^2+x_t^3}{4(1-x_t)^2}
-\frac{3x_t^3 \ln x_t}{2(1-x_t)^3}\,,
\nonumber\\
S_0(x_c.x_t) &=& x_c\left[\ln\frac{x_c}{x_t}-\frac{3x_t}{4(1-x_t)}
-\frac{3x_t^2\ln x_t}{4(1-x_t)^2}\right]\,,
\ea
With the same input as before, one obtains \cite{Herrlich}
\be
\quad\eta_1 =1.53\quad\eta_2 = 0.57 \quad \eta_3=0.47\,.
\ee

\subsection{Step III: Matrix elements}
\label{longdistance1}

Now
remember that the $C_i$ depend on $\mu_{OPE}$ and on the
definition of the $Q_i$ and the numerical change in the coefficients
due to the various choices possible is not negligible.
It is therefore important both from the phenomenological and fundamental
point of view that this dependence is correctly accounted for in
the evaluation of the matrix elements.
We can solve this in various ways.
\begin{itemize}
\item {\bf Stay in QCD} $\Rightarrow$ Lattice calculations.

\item{\bf QCD Sum Rules} This is using the method of Ref.~\mycite{SVZ} as reviewed
in these books by Colangelo and Khodjamirian.\cite{Khodjamirian}
\item {\bf Give up} $\Rightarrow$ Naive factorization.
\item {\bf Improved factorization}
\item {\bf  $X$-boson method} (or fictitious boson method)
\item {\bf  Large $N_c$} (in combination with something like
the $X$-boson method.) Here the difference is mainly in the
treatment of the low-energy hadronic physics. Three main approaches
exist of increasing sophistication.\footnote{Which of course means that
calculations exist only for simpler matrix-elements for the more sophisticated
approaches.}
\begin{itemize}
\item CHPT: As originally proposed by Bardeen-Buras-G\'erard \cite{BBG}
and now pursued mainly by Hambye and collaborators.\cite{Hambye}
\item ENJL (or extended Nambu-Jona-Lasinio model~\cite{ENJL}):
 As mainly done by
myself and J.~Prades.\citefive{BPBK}{BPscheme}{kptokpp}{BPdIhalf}{BPeps}
\item LMD or lowest meson dominance approach.\cite{LMD} These papers stay
with dimensional regularization throughout. The $X$-boson corrections
discussed below, show up here as part of the QCD corrections.
\end{itemize}
\item{\bf Dispersive methods} Some matrix elements can in principle be deduced
from experimental spectral functions.
\end{itemize}
Notice that there other approaches as well, e.g. the chiral quark
model.\citethree{PichdeRafael}{Bertolini}{Bochum} 
These have no underlying arguments why the $\mu$-dependence
should cancel, but as mentioned below, the importance of some effects
was first discussed in this context. I will also not treat the
calculations done using bag models and potential models which
similarly do not address the $\mu$-dependence issue.

\section{The Matrix elements: $X$-boson and other approaches}
\label{longdistance}

In this section I discuss the approaches summarized
in Sect. \ref{longdistance1}.
First I discuss the various approaches mainly in the framework
of $B_K$ and will later quote some results for the other quantities
as well, concentrating on the more recent analytical methods with an
emphasis on the work I have been involved in myself.

\subsection{Factorization and/or vacuum-insertion-approximation}

This is quite similar to the naive estimate for $K\to\pi\pi$ described
above except it is applied to the four-quark operator rather than to
pure $W$-exchange. So we use
\be
\langle0|\bar s_\alpha\gamma_\mu d_\alpha|K^0\rangle =
i\sqrt{2}F_K p_K
\ee
to get
\be
\langle \overline K^0|{\cal H}_{eff}|K^0\rangle = C_{\Delta S=2}(\mu)
\frac{16}{3}F_K^2 m_K^2\,.
\ee
Now the other results are usually quoted in terms of this one, the ratio
is the so-called bag-parameter~\footnote{Named after one of the early
models in which they were estimated.} $\hat B_K$.

So vacuum-insertion or
factorization yields
\be
\left.\hat{B}_K\right|_{\mbox{\small factorization}}\equiv 1\,. 
\ee
Using large $N_c$,
the part where the quarks of one $K^0$ come from two different
currents in ${\cal H}_{\mbox{eff}}$ has to be excluded yields
\be
\left.\hat{B}_K\right|_{\mbox{\small large $N_c$}}\equiv 3/4\,.
\ee

\subsection{Lattice Calculations}

There are many difficulties
associated with this approach at present. Recent reviews
are in Ref. \mycite{Sachrajda} where further references can be found.
A major breakthrough was realized in Ref. \mycite{LuscherLellouch}
where the problem of the contamination by other states than the
wanted $\pi\pi$ final state at the kaon~\cite{MaianiTesta} mass was solved.

\subsection{Sum Rule Calculations}

The method of sum rules have been used in several ways.
One can use the simple method of calculating spectral functions
in the presence of the weak interactions. This approach has been
used by many authors to calculate $B_K$. The original calculation  is
Ref.~\mycite{Chetyrkin}. Some others are~Ref.~\mycite{BKsumother}
These sum rules share the problem of a correct chiral limit behaviour
with many other sum rules that attempt to predict properties
of the pseudoscalars. 
In the case of $K\to\pi\pi$ decays the situation is even more difficult.
The underlying three point Green function with one extra insertion of the
weak operator is calculable in perturbative QCD but the problem is that
it is very difficult to extract the matrix elements from sum rules here.
The socalled~\cite{SVZ} phenomenological part of the sum rule
contains contributions from many intermediate states and there is
no dominance of the part where all three external legs are resonances.

This problem prompted a search for two-point sum rules that could be used
to determine the matrix elements. The approach developed by Pich, de Rafael
and collaborators~\cite{PichdeRafael2}
was to calculate the spectral function of the
effective Hamiltonian directly which should then be matched onto
the correct chiral form for the effective Hamiltonian at low energies
on the phenomenological side using a chirally correct form of resonance
saturation. This approach seemed to reproduce nicely the $K\to3\pi$
$\Delta I=3/2$ amplitude but gave a rather low value for $B_K$ and
could not reproduce the $\Delta I=1/2$ part of the amplitude.
The reason for the latter was found in Ref.~\mycite{PichdeRafael}, the
octet part of the sum rule turned out to have very large QCD corrections.
The reason for the low value of $B_K$ compared to other approaches
was uncovered in Ref.~\mycite{BPBK}. Away from the chiral limit, there are
low-energy operators that contribute to the sum rule but not to
the wanted matrix element.

\subsection{Improved Factorization}

This corresponds to first taking a matrix-element between a particular
quark and gluon external state of
${\cal H}_{\mbox{eff}}$. This removes the scheme and scale
dependence but introduces a dependence on the particular
quark external state chosen.
This can be found in the paper by Buras, Jamin and Weisz quoted
in Ref.~\mycite{two-loop} and has been extensively used by
H.-Y.~Cheng.\cite{cheng}
This yields a correction factor of
\be
\label{VIAimproved}
\left(1+r_1\frac{\alpha_S(\mu)}{\pi}\right)\quad r_1=-\frac{7}{6}
\ee
for $B_K$ and a 10 by 10
correction matrix for the $\Delta S=1$ case taking the place
of $r_1$ in Eq. (\ref{VIAimproved}).

\subsection{The $X$-boson method: a simpler case first.}

\begin{figure}
\centerline{\includegraphics[width=0.5\textwidth]{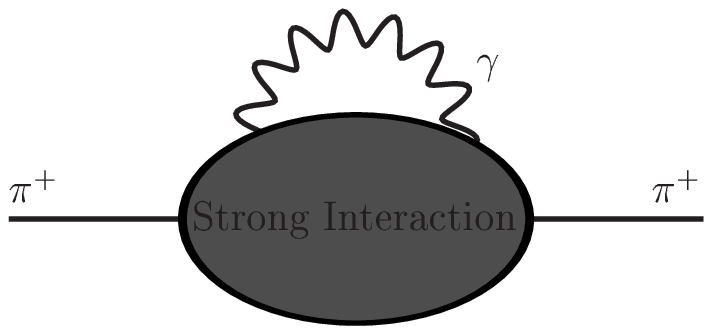}}
\caption{The electromagnetic contribution to the $\pi^+$-$\pi^0$
mass difference.}
\label{figpippio}
\end{figure}

Let us look at the simpler example of the electromagnetic
contribution to
$m_{\pi^+}^2-m_{\pi^0}^2$ in the chiral limit. This contribution
comes from one-photon exchange as depicted in Fig. \ref{figpippio}.

\begin{figure}
\centerline{\includegraphics[width=0.5\textwidth]{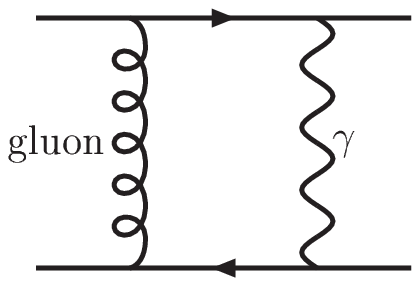}}
\caption{The short-distance photon-gluon box diagram leading to a four-quark
operator.}
\label{figboxphoton}
\end{figure}
The matrix element involves an integral over all photon momenta
\be
M=\dsp\int_0^\infty dq_\gamma^2\,.
\ee
We now split the integral at the arbitrary scale $\mu^2$.
The short-distance part of the integral,
$\dsp\int_{\mu^2}^\infty  dq_\gamma^2$,
can be evaluated
using OPE techniques \cite{BBG2} via the box diagram
of Fig. \ref{figboxphoton}. Other types of contributions are suppressed
by extra factors of $1/\mu^2$. The resulting four-quark operator
$\dsp (\bar{q}q)(\bar{q}q)$ can be estimated in large $N_c$\cite{BBG2}
\be
\left.  m_{\pi^+}^2-m_{\pi^0}^2 \right|_{\mbox{SD}} =
\frac{3\alpha_S\alpha_e}{\mu^2 F^4}\langle\bar {q}q\rangle^2\,.
\ee
The long-distance contribution,$\dsp\int_0^{\mu^2}dq_\gamma^2$,
can be evaluated in several ways.
CHPT at order $p^2$ or $p^4$ \cite{BBG2}, a vector meson dominance model(VMD)
\cite{BBG2}, the ENJL model~\cite{BPelem}
or LMD.\cite{LMD}
\begin{figure}[t]
\centerline{\includegraphics[width=0.8\textwidth]{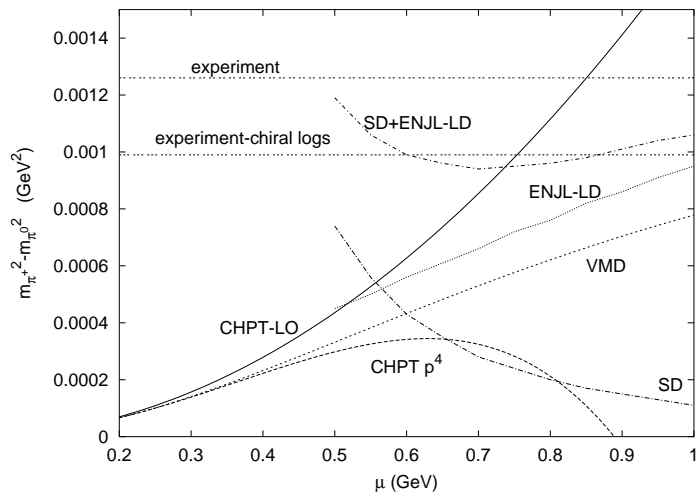}}
\caption{The Short-distance contribution (SD) and the various
versions of the long-distance contributions to
$m_{\pi^2}-m_{\pi^0}^2$. Also shown are the experimental value
and the experimental value minus the chiral logarithms that are extra.}
\label{figpiresult}
\end{figure}
The results are shown in Fig. \ref{figpiresult}. Notice that the sum
of long- and short- distance contributions is quite stable in the regime
$\mu\approx 500~$MeV to 1~GeV. VMD and LMD are the same in this case.

The main comments to be remembered are:
\begin{itemize}
\item The photon couplings are known {\em everywhere}.
\item We have a good identification of the scale $\mu$. It can
be identified from the photon momentum which is unambiguous.
\item In the end we got good matching, $\mu$-independence,
 and the numbers obtained agreed (maybe too) well with the experimental
result.
\end{itemize}

\subsection{The $X$-boson method.}

The improved factorization model is scheme- and scale-independent but
depends on the particular choice of quark/gluon state.
Now, photons are identifiable across theory boundaries, or
more generally, currents are~\footnote{At least the problem of
matching two-quark operators across theories is much more
tractable than four-quark operators.}. An example of this is CHPT
where the currents are the same as in QCD.

We can now try to get our four-quark operators back into something resembling
a photon so we can use the same method as in the previous section.
The full description including all formulas can be found in Ref.~\mycite{BPscheme}.
Similar work has been done by Bardeen.\cite{Bardeen}
For $\hat B_K$ this can be done by replacing
\be
{\cal H}_{\mbox{eff}}^{\Delta S=2}
\quad\mbox{ by }\quad
g_{X} X_{\Delta S=2}^\mu \left(\bar s_\alpha\gamma_\mu
 d_\alpha\right)_L\,.
\ee
with $M_X$ chosen such that $\alpha_S\log\frac{M_X}{\mu}$ is small and
we can neglect higher orders in $\mu^2/M_X^2$.

\begin{figure}
\includegraphics[width=\textwidth]{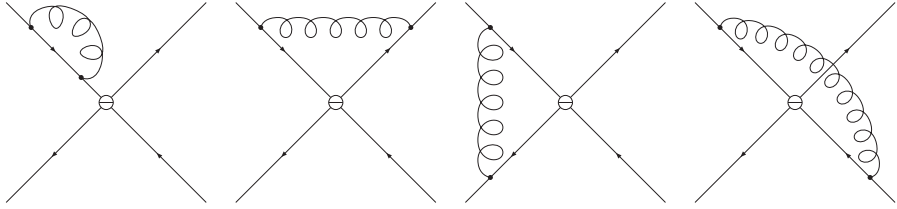}
\caption{The diagrams for the matrix-element of ${\cal H}_{\mbox{eff}}$
at one-loop.}
\label{figdiagOPE}
\end{figure}

We now take the matrix element of ${\cal H}_{\mbox{eff}}$ between
quark and gluon external states which yields from the diagrams
in Fig. \ref{figdiagOPE}
\begin{eqnarray}
&\dsp
iC_D\left[\left(1+\alpha_S(\nu)F(q_i)\right)S_1 
+\left(1+\alpha_S(\nu)F^\prime(q_i)\right)S_2\right]
&\nonumber\\&
C_D = -C(\nu)\left(1+\frac{\alpha_S(\nu)}{\pi}\left[\frac{\gamma_1}{2}
\ln\left(\frac{2q_1\cdot q_2}{\nu^2}\right)+r_1\right]\right)
&
\end{eqnarray}
with
$S_1$ and $S_2$ the tree level matrix elements between quarks of
$(\bar{s}\gamma^\mu d)_L
(\bar{s}\gamma_\mu d)_L$.

\begin{figure}
\includegraphics[width=\textwidth]{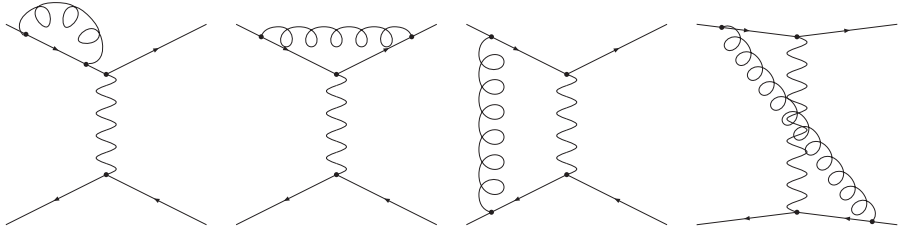}
\caption{The same matrix element but now of $X$-boson exchange.
The wiggly line is the $X$-boson.}
\label{figdiagX}
\end{figure}

We then calculate the same matrix element using $X$-boson exchange from
the diagrams in Fig. \ref{figdiagX}
and get
\ba
&\dsp\hskip-1.5cm
iC_C\left[\left(1+\alpha_S(\mu_C)F(q_i)\right)S_1 +
(1+\alpha_S(\mu_C)F^\prime(q_i))S_2\right]
+{\cal O}(M_X^{-4})
\nonumber\\
&
C_C = \frac{-g_X^2}{M_X^2}
\left(1+\frac{\alpha_S(\mu_C)}{\pi}\left[\frac{\gamma_1}{2}
\ln\left(\frac{2q_1\cdot q_2}{M_X^2}\right)+\tilde{r}_1\right]\right)
\ea

Notice that all the dependence on the external quark/gluon state
in the functions
$F(q_i)$ and $F^\prime(q_i)$ cancels.
 $r_1$ removes the 
scheme dependence and $\tilde{r}_1$ changes to the $X$-boson current scheme.

$g_X$ is now scale, scheme and external quark-gluon state independent.
It still depends on the precise scheme used for the vector and axial-vector
current. 

The
$\Delta S=1$ case is more complicated, everything becomes
10 by 10 matrices but can be found in Ref.~\mycite{BPeps}.
The precise definition of the total number of $X$-bosons needed
to discuss this case is
\ba
&\dsp
g_1 X_1^\mu \left((\bar s\gamma_\mu d)_L + (\bar u\gamma_\mu u)_L\right)
+ g_2 X_2^\mu  \left((\bar s\gamma_\mu u)_L + (\bar u\gamma_\mu d)_L\right)
&\nonumber\\
&\dsp +g_3 X_3^\mu \left((\bar s\gamma_\mu d)_L + 
\sum_{q=u,d,s}(\bar q\gamma_\mu q)_L\right)
+g_4 \sum_{q=u,d,s} X_{q,4}^\mu 
    \left((\bar s\gamma_\mu q)_L+(\bar q\gamma_\mu d)_L\right)
&\nonumber\\
&\dsp
+g_5 X_5^\mu \left((\bar s\gamma_\mu d)_L + 
\sum_{q=u,d,s}(\bar q\gamma_\mu q)_R\right)
+g_6 \sum_{q=u,d,s} X_{q,6} \left((\bar s  q)_L + 
(-2)(\bar q d)_R\right)
&\nonumber\\
&\dsp
+g_7 X_7^\mu \left((\bar s\gamma_\mu d)_L + 
\sum_{q=u,d,s}\frac{3}{2}e_q(\bar q\gamma_\mu q)_R\right)
&\nonumber\\&\dsp
+g_8 \sum_{q=u,d,s} X_{q,8} \left((\bar s  q)_L + 
(-2)\frac{3}{2}e_q(\bar q d)_R\right)
&\nonumber\\
&\dsp
+g_9 X_9^\mu \left((\bar s\gamma_\mu d)_L + 
\sum_{q=u,d,s}\frac{3}{2}e_q(\bar q\gamma_\mu q)_L\right)
&\nonumber\\
&\dsp
+g_{10} \sum_{q=u,d,s} X_{q,10}^\mu 
    \left((\bar s\gamma_\mu q)_L+
\frac{3}{2}e_q(\bar q\gamma_\mu d)_L\right).
&
\ea
The resulting change from this correction is displayed in columns 4 and 5
of Table \ref{tabzimu}. The corrections are substantial and turn
out to be in the wanted direction in all cases, surprisingly enough.

So let us summarize the $X$-boson scheme
\begin{enumerate}
\item Introduce a set of fictitious gauge bosons: $X$
\item $\alpha_S\log({M_X}/{\mu})$ does not need resumming, this is not large.
\item $X$-bosons must be {\em uncolored}.
\item Only perturbative QCD and OPE have been used so far.
\item For $\hat B_K$ we need $r_1-\tilde r_1 = -\frac{11}{12}$.
\end{enumerate}

\subsection{$X$-boson scheme matrix element: $\hat B_K$}

We now need to calculate
$\langle \mbox{out}|X\mbox{-exchange}|\mbox{in}\rangle$. First we
do the same split in the $X$-boson momentum integral as we did
for the photon
\be
{\int d q_X^2}\quad \Longrightarrow\quad
{\int_0^{\mu^2} d q_X^2+\int_{\mu^2}^\infty d q_X^2}
\ee
For $q^2_X$ large, the
kaon-form-factor suppresses direct mesonic contributions by $1/q_X^2$.
Large $q_X^2$ must thus flow back via {\em quarks-gluons}. The results
are already suppressed by $1/N_c$ so we can use leading $1/N_c$ in this part.
This part ends up replacing
$\dsp\log\frac{\mu_{OPE}}{M_X}$ by $\log\frac{\mu_{OPE}}{\mu}$ such that,
as it should be, the dependence on
$M_X$ has disappeared completely.

For the small $q^2_X$ integral we now
successively use better approximations in 3 directions:
\begin{itemize}
\item Low-energy that better approximates perturbative QCD
\item Inclusion of quark-masses
\item Inclusion of electromagnetism
\end{itemize}
The last step at present everyone only does at
short-distance.
One of the problems in the calculations is that 
Chiral Symmetry provides very strong constraints, which
lead to large cancellations between different parts. It is therefore
important that all contributions are calculated with a similar scheme
to take care of these cancellations correctly.

A few comments are appropriate:
\begin{itemize}
\item
The Chiral Quark Model approach \cite{Bertolini} does not do the
identification of scales and we do not include their results.
But they stressed large effects from {\em FSI, quark-masses} when
factorization+small variations was the main method. See also
Ref.~\mycite{PP}.
\item
For some matrix-elements CHPT allows to relate them to integrals
over measurable spectral functions.
The remainder agrees numerically for these $B_7,m_{\pi^+}^2-m_{\pi^0}^2$.
This is discussed in more detail in Sect.~\ref{dispersive}.
\end{itemize}

\begin{figure}
\centerline{\includegraphics[height=0.8\textwidth,angle=270]{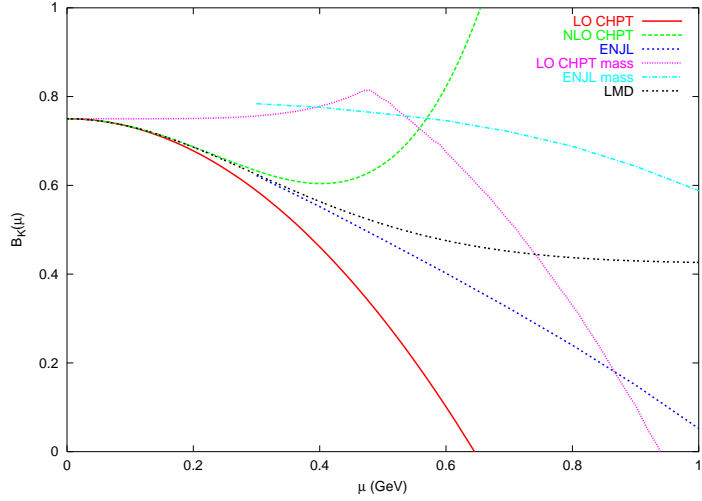}}
\caption{Comparison of the long-distance contributions 
to $B_K$ in the
various approximations discussed in the text.}
\label{figBKlong}
\end{figure}

The different results for
$B_K(\mu)$ in the chiral limit are
\ba
B_K^\chi(\mu) &=&
\frac{3}{4}\left[
1\mbox{ (large-$N_c$)}
-\frac{3\mu^2}{16\pi^2F_0^2}\mbox{ ($p^2$)}
\right.
\nonumber\\&&
\left.
+\frac{6\mu^4}{16\pi^2F_0^4}
\left(2L_1+5L_2+L_3+L_9\right)\mbox{ ($p^4$)}\right]
\ea
for CHPT~\citetwo{BPBK}{BPscheme}. The ENJL model
we do numerically~\citetwo{BPBK}{BPscheme}
and LMD gives~\footnote{
I have pulled factors of $\mu^2_{had}$ 
into the $\alpha_i,\beta_i$. Here and in the published version only
up to double poles are included. Triple poles were included
in the revised web version. They do improve the matching
over what is shown here.}~\cite{LMD}
\ba
B_K^\chi(\mu) &=&
\frac{3}{4}\left\{1-\frac{1}{32\pi^2F_0^2}\right.
\int_0^{\mu^2} dQ^2 
\times\nonumber\\&&
\left.\left(6-\sum_{i=res}\left[
\frac{\alpha_i}{Q^2+M_i^2}-\frac{\alpha_i}{M_i^2}
+\frac{\beta_i}{(Q^2+M_i^2)^2}-\frac{\beta_i}{M_i^4}\right]\right)\right\}
\ea
$\alpha_i$ and $\beta_i$ are particular combinations of the resonance
couplings. These we can now restrict by
comparing CHPT and LMD,
\be
\sum_i\frac{\alpha_i}{M_i^4}+\frac{\beta_i}{M_i^6}
=\frac{24}{F_0^2}\left(2L_1+5L_2+L_3+L_9\right)\,,
\ee
and using the short-distance constraints:
\ba
\dsp\sum_i \left(\alpha_i M_i^2 -\beta_i\right) &=& 0
\nonumber\\
\sum_i
\left(\frac{\alpha_i}{M_i^2}+\frac{\beta_i}{M_i^4}\right)&=&6
\nonumber\\
\sum_i \alpha_i &=& 24\pi^2\frac{\alpha_S}{\pi} F_0^2\,.
\ea
The last requirement is from explicitly requiring matching.
The various long distance contributions in the chiral limit and
in the presence of masses are shown in Fig. \ref{figBKlong}.

\begin{figure}
\centerline{\includegraphics[width=0.8\textwidth]{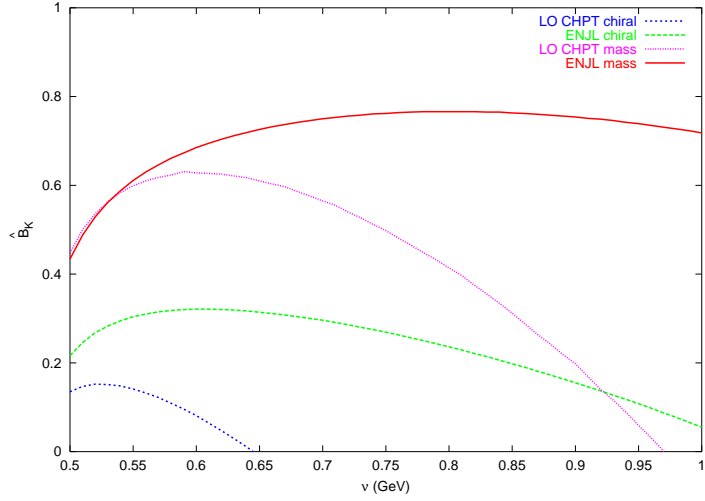}}
\caption{Using the long-distance depicted in Fig. \ref{figBKlong}
we obtain as results for $\hat B_K$ as a function of $\mu$.}
\label{figBK}
\end{figure}
Including the short-distance part leads to the results for $\hat B_K$
shown in Fig. \ref{figBK} and
\ba
\hat B_K^\chi &=& 0.32\pm0.06~(\alpha_S)\pm0.12~(\mbox{model})\nonumber\\
\hat B_K &=& 0.77\pm0.05~(\alpha_S)\pm0.05~(\mbox{model})\,.
\ea
The LMD model leads to somewhat higher 
but compatible results for the chiral case.\cite{LMD}
The results just using CHPT are also very similar but have worse
matching.\cite{BBG} The original results of Ref.~\mycite{BBG} have been
corrected for a correct momentum identification
in Ref.~\mycite{BGK} and Ref.~\mycite{BPscheme}.

\subsection{$X$-boson method results for $\Delta I=1/2$ rule
and $\varepsilon^\prime/\varepsilon$.}

We now present the results of the $X$-boson method also for
the $\Delta S=1$ quantities. For other approaches I refer to the
original references and various talks.\cite{Osaka}

The notation used below and more extensive discussions can be found
in Ref.~\mycite{BPeps}.

The lowest-order CHPT Lagrangian for the non-leptonic
$\Delta S=1$ sector is given by
\begin{eqnarray}
\label{CHPTdS1}
{\cal L}_{\Delta S=1} &=&
-CF_0^4\, \left[
{ G_8}\, \mbox{tr}(\Delta_{32}u_\mu u^\mu )
+ \, {G_8^\prime}\, \mbox{tr}(\Delta_{32}\chi_+ )
\right.
\nonumber \\&& 
+  {G_{27}} \, 
t^{ijkl}\mbox{tr}(\Delta_{ij} u_\mu u^\mu )
\mbox{tr}(\Delta_{kl}u_\mu u^\mu )
\left.+{e^2 G_E} F_0^2\mbox{tr}(\Delta_{32}\tilde Q)
\right] 
 ; 
\end{eqnarray}
and contains four couplings,
The various notations used are
\ba
U    \equiv 
\exp\left({i\sqrt{2} \Phi/F_0}\right) \equiv u^2 \, ;&&
u_\mu \equiv i u^\dagger  (D_\mu U) u \, ;
\nonumber\\
 \chi_+   \equiv   
 2 B_0 \left ( u^\dagger {\cal M} u^\dagger + 
u {\cal M}^\dagger u \right) \,
&&
\Delta_{ij}= u\lambda_{ij}u^\dagger\,;
\nonumber\\
(\lambda_{ij})_{ab}=\delta_{ia}\delta_{jb}\,;
&&
\tilde Q = u^\dagger Q u\,;
\nonumber\\
C = \frac{3G_F}{5\sqrt{2}}V_{ud}V_{us}^*\,.
\ea
This notation is very similar
to the notation used by Leutwyler for CHPT in his chapter.

Fixing the parameters from $K\to\pi\pi$
allows to predict $K\to3\pi$ to about 30\%. This we discuss
in Sect.~\ref{sectCHPT}.

In the limit
$N_c \to \infty$ \& $e\to0$ the parameters become
\be
 G_8,\, G_{27}  \longrightarrow 1\,; \quad
 G_8',\, e^2 G_E\longrightarrow 0\,. 
\ee
The normalization of $G_8$ and $G_{27}$ was chosen in order to have this
simple limit.

The isospin 0 and 2 amplitudes for $K\to\pi\pi$ from the
above Lagrangian are
\ba
{a_0} & = &\frac{\sqrt 6}{9} \, C F_0\left[\left(9G_8+G_{27}\right) 
(m_K^2-m_\pi^2) - 6 e^2 G_E F_0^2 \right]
\nonumber\\
{a_2} & = & \frac{\sqrt 3}{9} \, C F_0\left[10 G_{27} \, 
(m_K^2-m_\pi^2) - 6 e^2 G_E F_0^2 \right]\,.
\ea
The experimental values are~\citetwo{KMW}{kptokpp}
Re$(G_8)\approx 6.2$ and Re$(G_{27})\approx0.48$ with a sizable error.

\begin{figure}
\begin{minipage}[b]{0.48\textwidth}
\includegraphics[height=0.99\textwidth,angle=270]{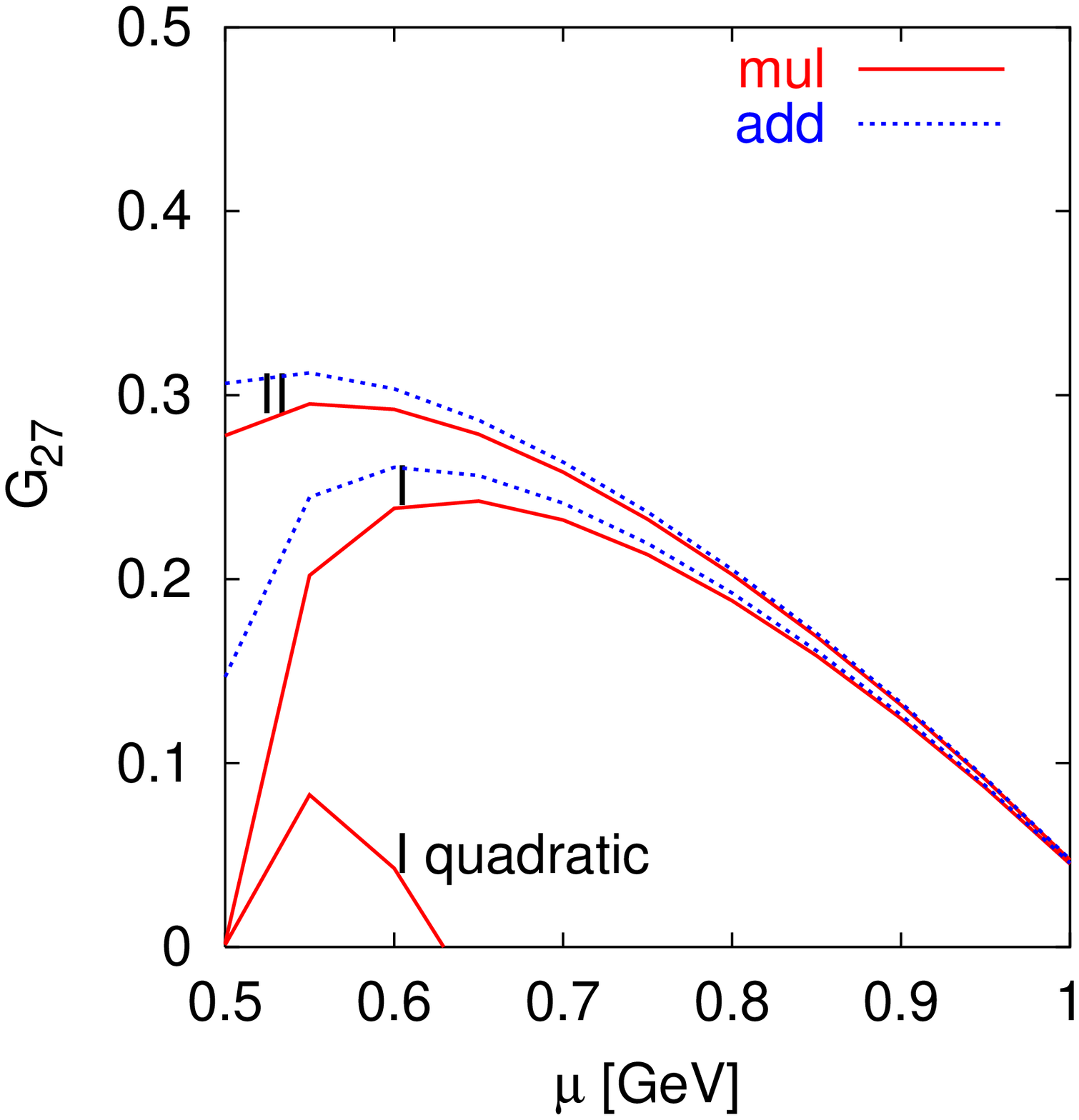}
\end{minipage}
\hspace{0.02\textwidth}
\begin{minipage}[b]{0.48\textwidth}
\includegraphics[height=0.99\textwidth,angle=270]{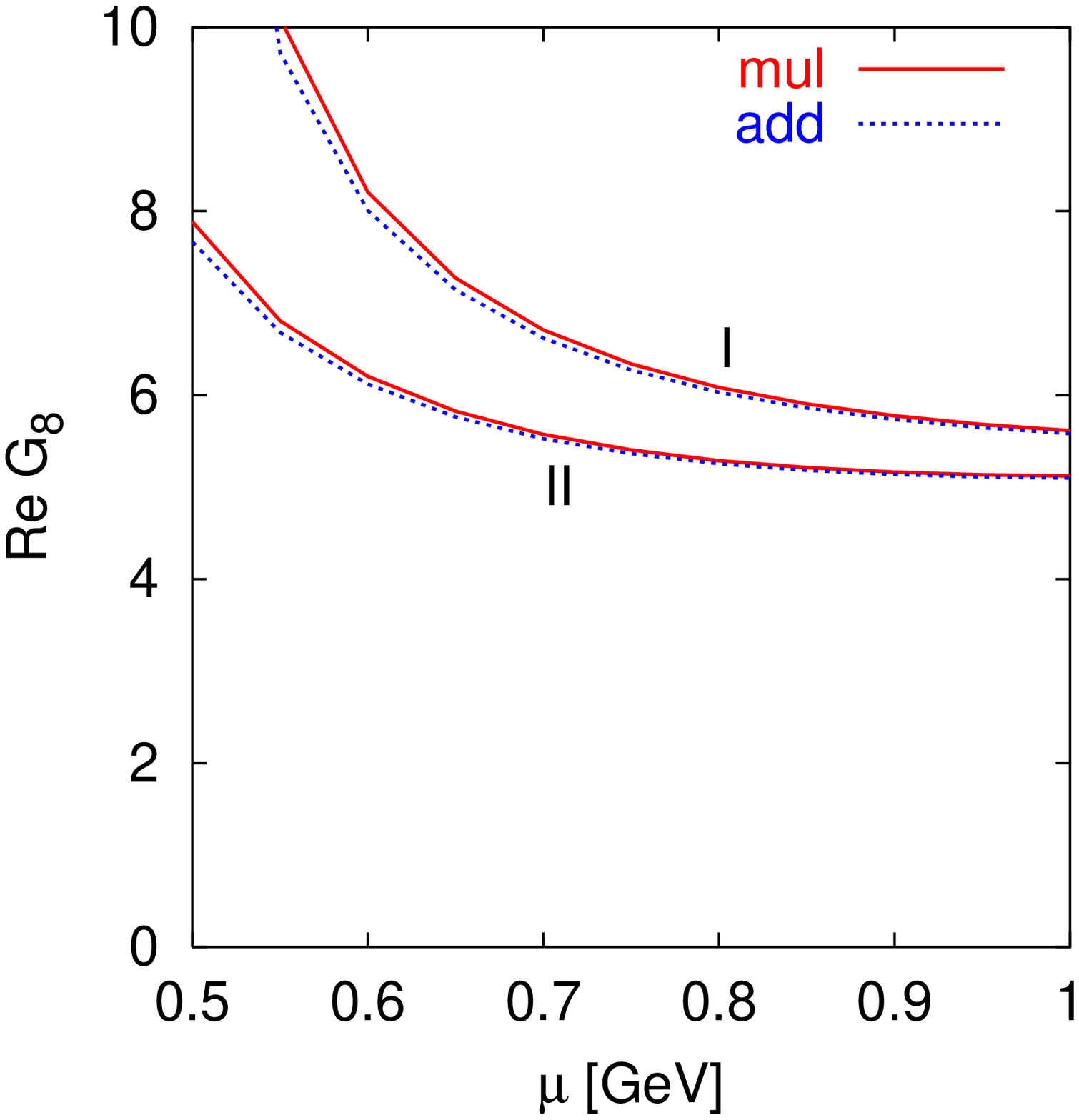}
\end{minipage}
\caption{Results for the real part of the $\Delta S=1$ chiral Lagrangian.
Remember that
$\re G_8^{\mbox{exp}}\approx 6.2$ and 
$\re G_{27}^{\mbox{exp}}\approx 0.48$.
The labels I and II refer to two different values of $\alpha_S$ and
mul-add to two ways of combining the various QCD corrections, differing only
at higher orders.
}
\label{figreal} 
\end{figure}
\begin{figure}
\begin{minipage}[b]{0.48\textwidth}
\includegraphics[height=0.98\textwidth,angle=270]{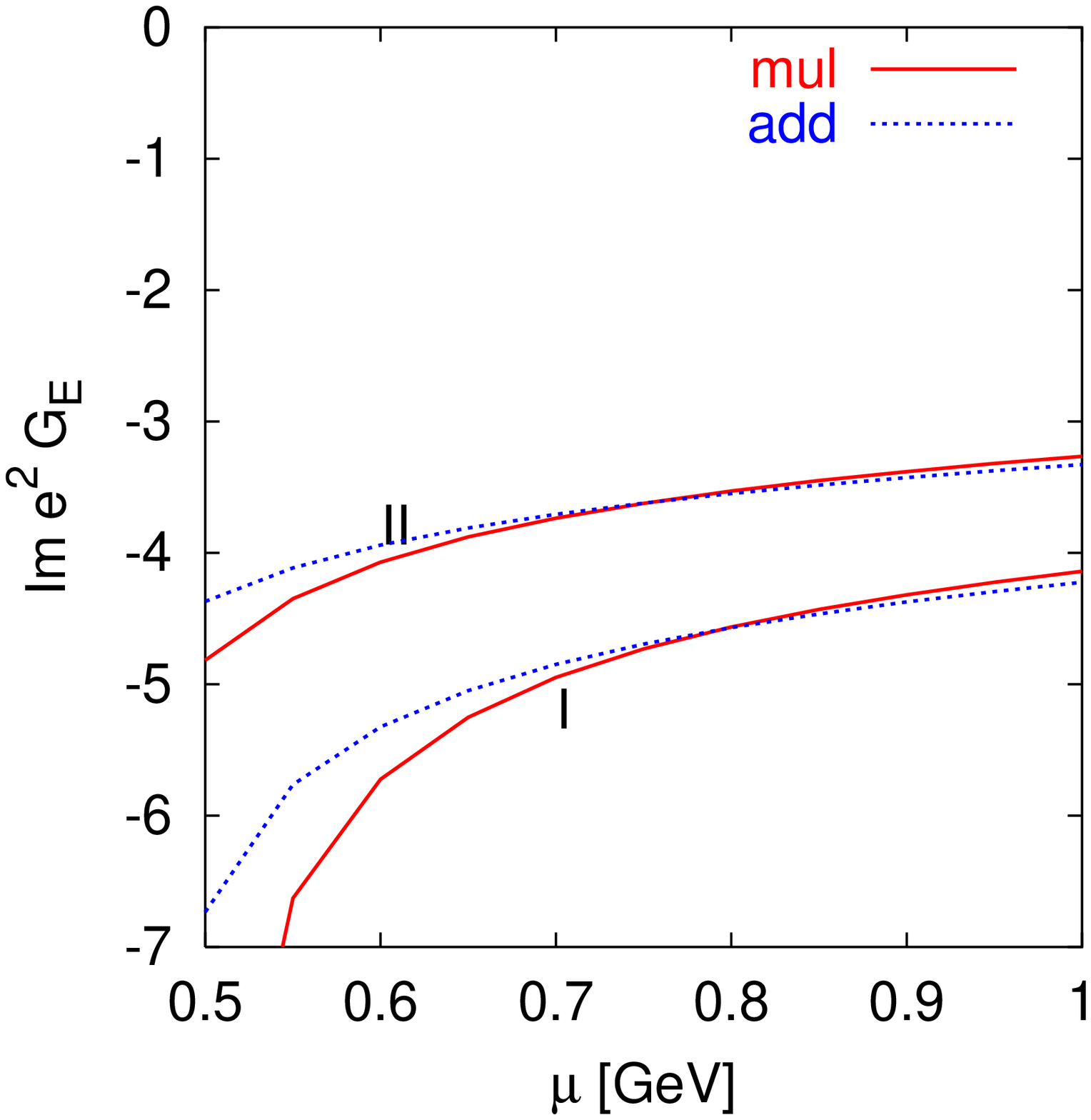}
\end{minipage}
\hspace{0.02\textwidth}
\begin{minipage}[b]{0.48\textwidth}
\includegraphics[height=0.98\textwidth,angle=270]{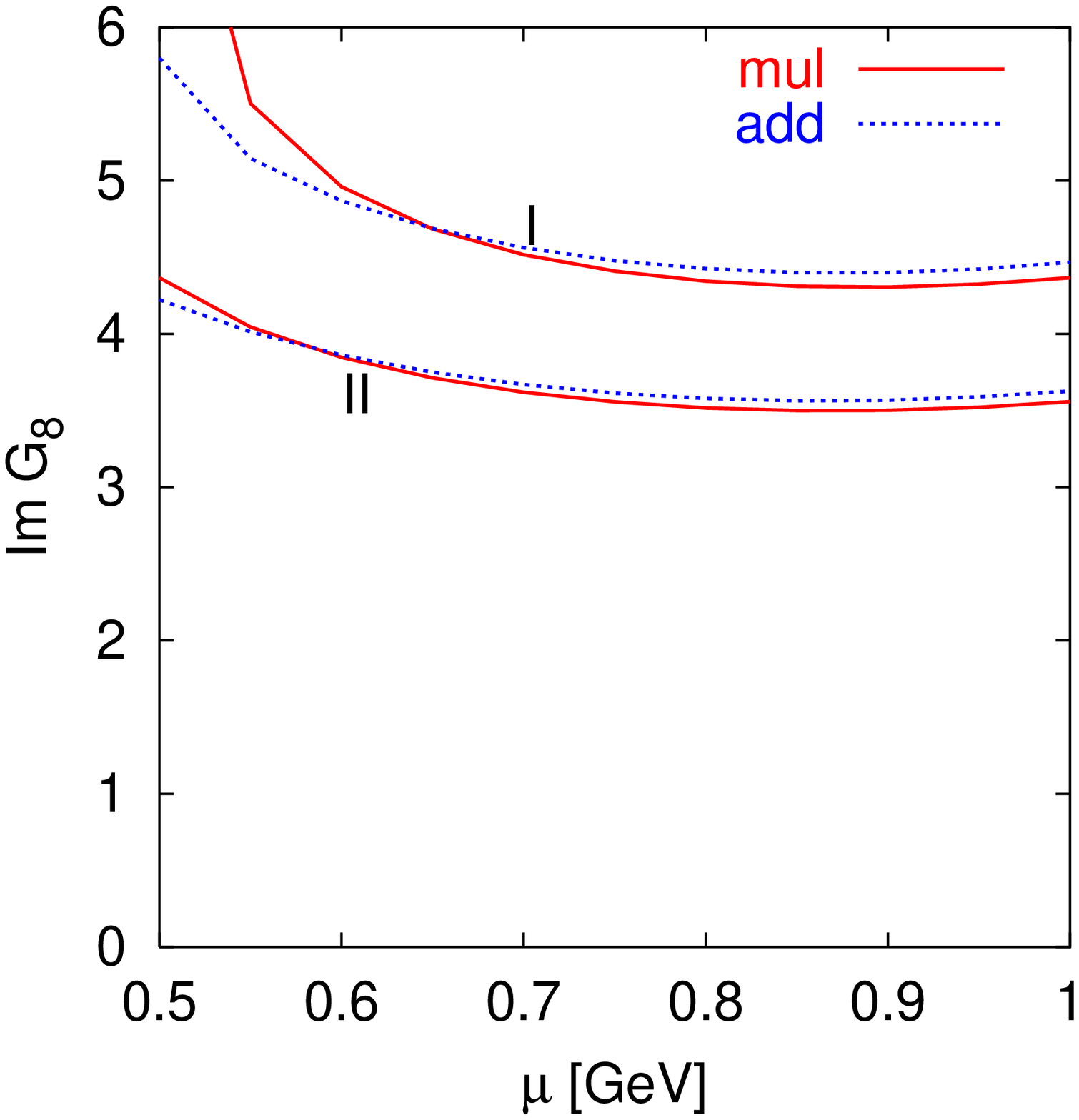}
\end{minipage}
\caption{Results for the imaginary part of the $\Delta S=1$ chiral Lagrangian.}
\label{figimag}
\end{figure}

The results obtained are shown in Fig. \ref{figreal} for the real
parts and in Fig. \ref{figimag} for the imaginary parts.
Notice that we get good matching for most quantities and good agreement
with the experimental result for $G_8$. The bad matching for
$G_{27}$ is because we have a large cancellation needed
between the non-factorizable and the factorizable case to obtain matching.
The  30\% or so accuracy we have on the non-factorizable part leads therefore
to large errors on the final result. Notice that
all methods that do not impose the matching by hand, typically have a
problem with this.  The other quantities are not affected
by such a cancellation and show therefore better matching.
The origin of the large enhancement for $G_8$ is in the long-distance Penguin
diagrams. The long distance calculations contains contributions that
have the same quark line structure of the gluonic Penguin diagram.
It is these that provide the extra enhancement. Otherwise we
would have the relation to next-to-leading order in $1/N_c$
of $1-G_{27} = G_8-1$.\cite{PRlargeNc}

\begin{figure}
\centerline{\includegraphics[height=0.8\textwidth,angle=270]{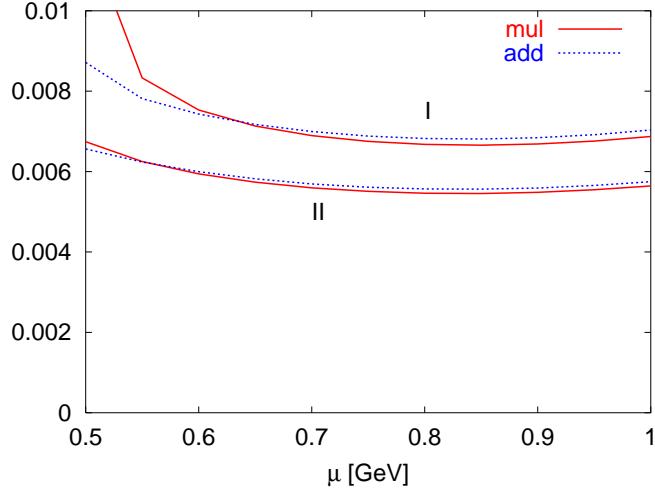}}
\caption{The results for $\varepsilon^\prime/\varepsilon$ in the chiral limit.
Notice the quite good matching.}
\label{figeps}
\end{figure}

We can now use the above results to estimate
$\varepsilon^\prime/\varepsilon$ in the chiral limit.
We used $G_{27}=0.48$, $\re G_8=6.2$ and the values we obtained for the
imaginary part. The same method leads to
$\varepsilon$ within 10\% of the experimental value. The result
is shown in Fig. \ref{figeps}.
We can conclude that
\be
\dsp\left(\frac{\varepsilon^\prime}{\varepsilon}\right)^\chi
= (7.4-1.9)\cdot10^{-3}=5.5\cdot10^{-3}
\ee
and\\
$\bullet$
 $B_6\approx 2.5$ not $\lesssim 1.5$\\
$\bullet$  $B_8\approx 1.3$ OK but not $B_8\approx B_6$\,.

Using
\be
\label{epsp}
|\varepsilon^\prime| \simeq
\frac{1}{\sqrt{2}}\frac{\re \, a_2}{\re \, a_0}
\left(-\frac{\im \, a_0}{\re \, a_0}+\frac{\im \, a_2}{\re \, a_2}\right)
\ee
We can now include the two main known corrections.
The usual approach for final state interactions (FSI)
is to take
$\re a_0$, $\re a_2$ from experiment and
$\im a_0$, $\im a_2$ to ${\cal O}(p^2)$. This leads
to a large suppression of the first term in Eq. (\ref{epsp}).\cite{PP}
We evaluate both to $p^2$ so for us
FSI act mainly on the  prefactor in Eq. (\ref{epsp}).

The main isospin breaking correction
is that
$\pi^0$,$\eta$ and $\eta^\prime$ mix. This, together with
the other isospin breaking corrections, brings in a part
of the large $a_0$ into $a_2$ and is thus enhanced. The
effect is usually parametrized as
\be
\frac{\Delta\im a_2}{\re a_2}\approx\Omega\frac{\im a_0}{\re a_0}
\quad\mbox{with}\quad \Omega\approx 0.16\pm0.03
\ee
where the numerical value is taken from Ref.~\mycite{EPetal}.
There is, in addition to Ref. \mycite{EPetal}
a rather large body of recent work on electromagnetic and isospin
corrections.\cite{EMnonleptonic}

Including the last two main corrections yields
\be
\left|\frac{\varepsilon^\prime}{\varepsilon}\right| = (5.4-2.3)\cdot10^{-3}
= (3.1\pm ??)\cdot10^{-3}\,.
\ee
The size of the error is debatable but should be at least 50\% given
all the uncertainties involved.

\subsection{Dispersive Work and Higher Order Operators}
\label{dispersive}

Some of the matrix elements we want here can be extracted from
experimental information in a different way. The canonical
example is the mass difference between the charged and the neutral pion
in the chiral limit which can be extracted from a dispersive integral
over the difference of the vector and axial vector spectral
functions.\cite{Dasetal}

This idea has been pursued in the context of weak decay in a series of
papers by Donoghue, Golowich and collaborators.\cite{DG1}
The matrix element of $Q_7$ could be extracted directly
from these data. To get at the matrix element of $Q_8$ is somewhat more
difficult. Ref. \mycite{DG1} extracted it first by requiring $\mu$-independence.
In the first paper cited in Ref.~\mycite{LMD}, it was realized that the matrix
element of $Q_8$ could also be extracted from the spectral functions
and was related to the coefficient of the dimension 6 term in the operator
product expansion of the underlying Green's function.
The most recent papers using this method are Refs.~\mycite{KnechtQ7Q8},
\mycite{NAR01},
\mycite{DG2} and \mycite{BPQ7Q8}.
In the last two papers also some QCD corrections were included which had a
substantial impact on the numerical results.

The results are given in Table.~\ref{tableQ7Q8}. The operator
$O_6^{(1)}$ is related by a chiral transformation to $Q_7$
and $O_6^{(2)}$ to $Q_8$.
The numbers are valid in the chiral limit.
\begin{table}
\begin{center}
\begin{tabular}{|c|c|c|}
\hline
Reference&$\langle 0| O_6^{(1)}|0 \rangle^{NDR}_\chi$&
$\langle 0| O_6^{(2)}|0 \rangle^{NDR}_\chi$ 
 \\ 
\hline
$B_7=B_8=1$                      & $-(5.4\pm2.2)\cdot10^{-5}$ GeV$^6$ & $(1.0 \pm 0.4)\cdot10^{-3}$ GeV$^6$   \\
\hline				   
Bijnens et al. \cite{BPQ7Q8}     &$-(4.0\pm0.5)\cdot10^{-5}$ GeV$^6$& $(1.2  \pm 0.5)\cdot10^{-3}$ GeV$^6$  \\  
Knecht et al. \cite{KnechtQ7Q8}  &$-(1.9\pm0.6)\cdot10^{-5}$ GeV$^6$& $(3.5 \pm 1.1)\cdot10^{-3}$ GeV$^6$   \\   
Cirigliano et al. \cite{DG2}     &$-(2.7\pm1.7)\cdot10^{-5}$ GeV$^6$& $(2.2 \pm 0.7)\cdot10^{-3}$ GeV$^6$   \\  
\hline				   
Donoghue et al.\cite{DG1}        &$-(4.3\pm0.9)\cdot10^{-5}$ GeV$^6$& $(1.5 \pm 0.4)\cdot10^{-3}$ GeV$^6$      \\  
Narison        \cite{NAR01}      &$-(3.5\pm1.0)\cdot10^{-5}$ GeV$^6$& $(1.5 \pm 0.3)\cdot10^{-3}$ GeV$^6$      \\  
lattice   \cite{Q7Q8lattice}     &$-(2.6\pm0.7)\cdot10^{-5}$ GeV$^6$& $(0.74 \pm 0.15)\cdot10^{-3}$ GeV$^6$    \\
ENJL  \cite{BPeps}               &$-(4.3\pm0.5)\cdot10^{-5}$ GeV$^6$& $(1.3\pm0.2)\cdot10^{-3}$ GeV$^6$\\
\hline
\end{tabular}
\end{center}
\caption{The values of the VEVs in the NDR scheme at $\mu_R=2$ GeV.
The most recent dispersive results are line 3 to 5.
Adapted from Ref.~[\ref{biblio:BPQ7Q8}].
\label{tableQ7Q8}}
\end{table}
The various results for the matrix element of $O_6^{(1)}$ are in reasonable
agreement with each other. The underlying spectral integral,
evaluated directly from data in Refs.~\mycite{NAR01},\mycite{DG2} and
 \mycite{BPQ7Q8},
or via the minimal hadronic ansatz~\cite{KnechtQ7Q8} are in
better agreement. The main difference is that the number quoted for
Ref. \mycite{KnechtQ7Q8} does not include the extra QCD corrections.
The largest source of the differences is the way the different
results for the underlying evaluation of $O_6^{(2)}$ come back into
$O_6^{(1)}$.

The results for $O_6^{(2)}$ differ more. Ref. \mycite{BPQ7Q8}
uses two approaches. First, the matrix element for $O_6^{(2)}$ can be extracted
via a similar dispersive integral over the scalar and pseudoscalar spectral
functions. The requirements of short-distance matching for this spectral
function combined with a saturation with a few states imposes that
the nonfactorizable part is suppressed and the number and error quoted follows
from this. Extracting the coefficient of the dimension 6 operator
in the expansion of the vector and axial-vector spectral functions yields a
result comparable but with a larger error of about 0.9.
Ref.~\mycite{DG2} chose to enforce all the known constraints on the
vector and axial-vector spectral functions to obtain a result. This resulted
in rather large cancellations between the various contributions making an error
analysis more difficult. A reasonable estimate lead to the value quoted.
Ref.~\mycite{KnechtQ7Q8} did not include the extra QCD correction which lowers
the value. The number quoted is derived by a single resonance plus continuum
ansatz for the spectral functions and assuming a typical large $N_c$
error of 30\%. This ansatz worked well for lower moments of the spectral
functions which can be tested experimentally. Adding more resonances allows
for a broader range of results.\cite{BPQ7Q8}
The reason why these numbers based on the same data can be so different is
that the quantity in question is sensitive to the energy regime
above 1.3~GeV where the accuracy of the data is rather low.

\section{CHPT tests in non-leptonic Kaon decays to pions}
\label{sectCHPT}

The use of current algebra methods in kaon decays to
pions goes back a long time.\cite{Cronin} In Ref.~\mycite{Cronin}
the full lowest order CHPT contributions were worked out for
kaon decays to pions using current algebra methods. Later it
was realized~\cite{DonoghueBK} that $B_K$ could be determined
from the $\Delta I=3/2$ part of $K \to\pi\pi$ and a substantial
deviation from the vacuum saturation value was found, about $0.3$
instead of $1$. It was found later that this particular 3-flavour chiral
symmetry relation has potentially large corrections~\cite{BSW}
and the full analysis of Ref.~\mycitetwo{KMW}{KMW2} showed that
there are free parameters in this relation. Nevertheless, CHPT is
still useful for nonleptonic kaon decays for connecting
$K\to\pi\pi$ to $K\to\pi\pi\pi$.

\begin{figure}
\centerline{\includegraphics[width=0.9\textwidth]{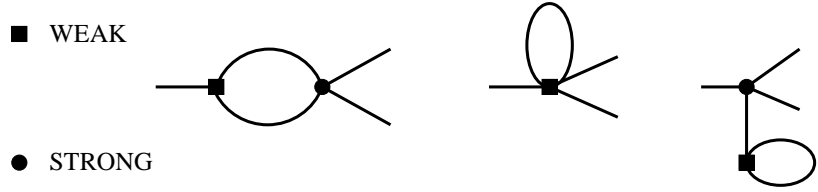}}
\caption{The CHPT diagrams with meson loops at order $p^4$
for $K\to\pi\pi$.}
\label{figk2pi}
\end{figure}

I have already shown the lowest order $\Delta S=1$ CHPT Lagrangian
in Eq. (\ref{CHPTdS1}) and mentioned that this
reproduces $K\to\pi\pi\pi$ to about 30\% from $K\to\pi\pi$.
This can be extended to an order $p^4$ calculation in
CHPT~\citetwo{KMW}{KMW2}. The chiral logarithms for $K\to\pi\pi$
were known earlier.\cite{Bijnens85}
The diagrams needed, now the lines are mesons, not quarks as in most
of the previous figures, are shown in Fig. \ref{figk2pi} for $K\to\pi\pi$.
The $K\to\pi\pi\pi$ ones are similar.

In terms of the number of parameters and observables we
have
\begin{tabbing}
\# parameters :~\= $p^2$ :~\= 2 {(1)} \= $G_8$, $G_{27}$\\
\> $p^4$ :\> 7{(3)} \>
($-2$ that cannot be disentangled \\ \>\>\> from $G_8$, $G_{27}$)\\[0.5cm]
\# observables:\> After isospin\\[0.2cm]
$K\to 2\pi$ :\> 2{(1)}\\
$K\to3\pi$ :\> 2{(1)}{+1}\>\>constant in Dalitz plot\\
\> 3{(1)}{+3}\>\>linear\\
\> 5{(1)}\>\>quadratic\\
\end{tabbing}
Phases (the {+i} above) of up to linear terms in the
Dalitz plot might also be measurable.
The numbers in brackets refer to $\Delta I=1/2$ parameters and observables
only. Notice that a significant number of tests is possible.
Comparison with the present data is shown in Table \ref{tabk3pi}.
The numbers in brackets refer to which inputs produce which predictions.
\begin{table}
\begin{center}
\begin{tabular}{|c|ccc|}
\hline
\mbox{variable} & $p^2$ & $p^4$ &\mbox{experiment} \\
\hline
$\alpha_1 $& 74 & (1)\mbox{input} & 91.71$\pm$0.32\\
$\beta_1  $& $-$16.5  & (2)\mbox{input} & $-$25.68$\pm$0.27\\
$\zeta_1  $& & (1)$-$0.47$\pm$0.18 & $-$0.47$\pm$0.15\\
$\xi_1    $& & (2) $-$1.58$\pm$0.19 & $-$1.51$\pm$0.30\\
$\alpha_3 $& $-$4.1 &(3) \mbox{input} & $-$7.36$\pm$0.47\\
$\beta_3  $& $-$1.0 &(4) \mbox{input} & $-$2.42$\pm$0.41\\
$\gamma_3 $& 1.8  &(5) \mbox{input} & 2.26$\pm$0.23\\
$\xi_3    $& 6 & (4) 0.92$\pm$0.030 & $-$0.12$\pm$0.17\\
$\xi_3^\prime$& &(5) $-$0.033$\pm$0.077 & $-$0.21$\pm$0.51\\
$\zeta_3  $& &(3) $-$0.0011$\pm$0.006 & $-$0.21$\pm$0.08\\
\hline
\end{tabular}
\end{center}
\caption{CHPT to order $p^4$ for $K\to\pi\pi\pi$. The variables
refer to various measurables in the Dalitz plot.
$K\to\pi\pi$ is always used as input. Numbers in brackets
indicate relations.}
\label{tabk3pi}
\end{table}
It is important that in the future experiments tests these relations directly.
At present there is satisfactory agreement with the data.
Notice that the new CPLEAR data decrease the errors somewhat. A recalculation
of this process together with a fit to the newer data is in
progress.\cite{BDP}

CP-violation in $K\to3\pi$ will be very difficult. The strong phases
 needed to interfere are just
too small (Ref.~\mycite{KMW2} last reference).
E.g. 
$\delta_2-\delta_1$ in $K_L\to\pi^+\pi^-\pi^0$ is expected to be $-0.083$
and the present experimental bound is only $-0.33\pm0.29$.
The CP-asymmetries expected are about $10^{-6}$
so we expect in the near future only to improve limits.
A review of $CP$ violation in $K\to3\pi$ can be found in Ref.~\mycite{DI}.

\section{Kaon rare decays}

The below is a summary of the summary by Isidori given at
KAON99 and LP01
and Buchalla in KAON2001.\cite{Isidori}
 I refer there for references. Another somewhat
older but more extensive review is Ref.~\mycite{Littenberg} and I also
found Ref.~\mycite{Buras2} useful.

Some of the processes mentioned below are tests of strong interaction physics,
often in the guise of CHPT, and others are mainly SM tests.

\begin{itemize}
\item
{$\mathbf K^+\to\pi^+\nu\bar\nu$,$ \mathbf K_L\to\pi^0\nu\bar\nu$}
In this case the SM is strongly suppressed and dominated by
short-distance physics. It is thus
ideal for precision SM CKM tests and
possibly new physics searches.
The reason is that real and imaginary part of the amplitude are
similar here in size, CKM angle suppression is counteracted by the
large top-quark mass. This allows it to be dominated 
by $\bar s d Z$-Penguin and $WW$-box diagrams.
The resulting
\be
{\cal H}_{\mbox{eff}} = C_\nu (\bar s \gamma_\mu d)_L(\bar\nu\gamma^\mu\nu)_L
\ee
can be hadronized using the measured matrix-element from $K_{\ell3}$
and the
$\bar\nu\nu$ pair is in a CP eigenstate allowing lots of CP-tests.
The main disadvantage is the extremely low predicted branching ratio
of
\ba
\mbox{Neutral mode: }&& (3.1\pm1.3)\cdot10^{-11}\nonumber\\
\mbox{Charged mode: }&& (8.2\pm3.2)\cdot10^{-11}
\ea
This process will be competitive with $B$-decays in next generation of
kaon experiments.
\item
{$\mathbf K_L\to\ell^+\ell^-$}: The short-distance contribution comes
from $Z$-penguin and boxes. The main uncertainty comes from the
long-distance 2$\gamma$ intermediate state.

$K_L\to \mu^+\mu^-$ dominated by unitary part of $K_L\to\gamma\gamma$,
which can be taken from the branching ratio for that
decay. It fits the data well.

The long distance part of
$K_L\to e^+ e^-$ is more dependent on the contributions with off-shell
photons. Here there is still work to do.
\item {$\mathbf K\to\pi\ell^+\ell^-$}
The
real parts can be predicted by CHPT at order $p^4$ from 2 parameters, it
fits well.
For the imaginary part there are 
problems with long-distance contributions from $K\to\pi\gamma\gamma$.
but the CP-violating quantities are often dominated by direct part.
\item {$\mathbf K_S\to\gamma\gamma$}
This process was a parameter-free prediction from CHPT
 at order $p^4$ from the diagrams in Fig. \ref{figkgg}.
\begin{figure}
\includegraphics[width=\textwidth]{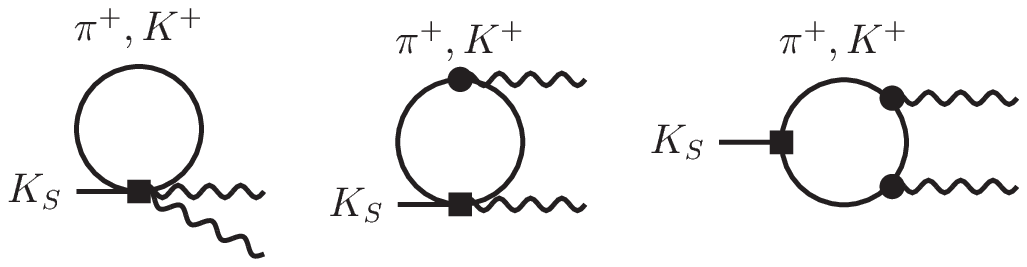}
\caption{The meson-loop diagrams contributing to $K_S\to\gamma\gamma$.
They predict the rate well.}
\label{figkgg}
\end{figure}

\begin{figure}
\centerline{\includegraphics[width=0.5\textwidth]{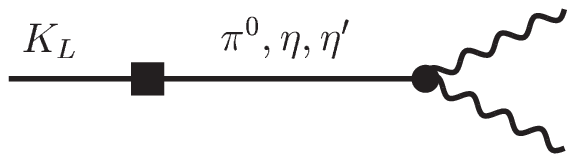}}
\caption{The main diagram for $K_L\to\gamma\gamma$ with a large
uncertainty due to cancellations.}
\label{figklgg}
\end{figure}
\item {$\mathbf K_L\to\gamma\gamma$}
This decay needs more work. The underlying difficulty is that
the main contribution is full of cancellations. The main diagram
is shown in Fig. \ref{figklgg}.
\item {$\mathbf K_L\to\pi^0\gamma\gamma$}
This process at $p^4$ is again a parameter-free CHPT prediction.
The spectrum is well described but the rate is somewhat off.
This can be explained by $p^6$ effects.
\item {$\mathbf K_S\to\pi^0\gamma\gamma$}
This process has very similar problems as $K_L\to\gamma\gamma$.
\item {$\mathbf K_{L(S)}\to\gamma^*\gamma^*$}
The same processes as above but with one or both photons off-shell, decaying
into a $\ell^+\ell^-$-pair.
These have similar questions/problems/successes as the ones with on-shell
photons.

\end{itemize}

\section{Conclusions} 

In this review I have given a historical overview and an
introduction of which sector of the standard model we hope to test
using these experiments. The main underlying problem is the
strong interaction and I have discussed their impact on the various
weak decays of light flavours. The results can be summarized as follows: 
\begin{itemize}
\item {\bf Semi-leptonic Decays}
\begin{itemize}
\item CHPT is a major success and tool here.
\item These decays are the main input for $V_{ud}$ and $V_{us}$
\item In addition they provide several tests of strong interaction effects.
\end{itemize}

\item {\bf $\mathbf K\to\pi\pi$ and $\mathbf \kob$-$\mathbf K^0$ mixing}
This was the main part of this review.
I hope I have convinced you that there is qualitative agreement
and successful prediction is in principle possible but more
work on including extra effects and pushing down the uncertainty is 
obviously needed.
\item{$\mathbf K\to\pi\pi\pi$} A good test of CHPT and strong interaction
effects but not very promising for CP violation studies.
\item{\bf Rare Decays}. I only presented a very short
summary of the issues.
\end{itemize}

\section*{Acknowledgements}
This work has been partially supported  by the Swedish Research Council
and by the European Union TMR Network
EURODAPHNE (Contract No. ERBFMX-CT98-0169).

\end{document}